\documentclass[showpacs,aps,pre]{revtex4}
\usepackage{amsmath}
\usepackage{graphicx}
\usepackage{bbm}
\include{epsf}
\usepackage{fancyhdr}
\usepackage{amssymb}

\def\(({\left(}
\def\)){\right)}

\newcommand{\be}{\begin{equation}}
\newcommand{\ee}{\end{equation}}

\newcommand{\da}{\partial a}
\newcommand{\di}{\partial i}
\newcommand{\djj}{\partial j}
\newcommand{\dk}{\partial k}

\newcommand{\ef}[1]{\, #1}     
\newcommand{\dx}[1] {\mathrm{d}{#1}}

\renewcommand{\setminus}{\smallsetminus}

\newcommand{\eval}[1]{\left\langle {#1} \right\rangle} 
\newcommand{\deenne}[2]{\frac{\partial^#2}{\partial #1 ^#2}} 
\newcommand{\smfrac}[2]{\genfrac{}{}{0.25pt}{1}{#1}{#2}}
\newcommand{\ham}{\mathcal{H}}
\newcommand{\ngg}[1]{\bar{#1}}

\begin{document}

\title{The Phase Diagram of 1-in-3 Satisfiability Problem}

\author{Jack Raymond$^1$, Andrea Sportiello$^{2}$, Lenka Zdeborov\'a$^3$}

\affiliation{%
$^1$ NCRG; Aston University, Aston Triangle, Birmingham, B4  7EJ \\
$^2$ Universit\`a degli Studi di Milano, via Celoria 16, I-20133 Milano \\
$^3$ CNRS; Univ.~Paris-Sud, UMR 8626, Orsay CEDEX, France 91405, LPTMS%
}

\begin{abstract}
We study the typical case properties of the 1-in-3 satisfiability
problem, the boolean satisfaction problem where a clause is satisfied
by exactly one literal, in an enlarged random ensemble parametrized by
average connectivity and probability of negation of a variable in a
clause. Random 1-in-3 Satisfiability and Exact 3-Cover are special cases of this
ensemble.  We interpolate between these cases from a region where
satisfiability can be typically decided for all connectivities in
polynomial time to a region where deciding satisfiability is hard, in
some interval of connectivities.  We derive several rigorous results
in the first region, and develop the one-step--replica-symmetry-breaking 
cavity analysis in the second one. We discuss the prediction 
for the transition between the almost surely satisfiable and the almost surely 
unsatisfiable phase, and other structural properties of the phase diagram, 
in light of cavity method results.
\end{abstract}

\date{\today}

\maketitle

\section{Introduction and motivation}

Classification of the average-case computational complexity of the
constraint satisfaction problems is a major task in theoretical
computer science.  Many problems were successfully analyzed by
rigorous probabilistic methods.  However, the average-case complexity
remains an open question for most of the well known NP-complete
problems \cite{cook, GJ}, for example the $K$-satisfiability, Vertex
Coloring, and also 1-in-$K$ satisfiability, on the commonly studied
random ensembles (sparse random regular or Erd\"os-R\'enyi graphs).

In recent years heuristic methods from statistical physics 
\cite{MP99, MP03, BMW00} have allowed us to understand some average-case properties 
of large random instances \cite{review}. The aim of these studies was not to prove
average NP-completeness \cite{Lev86}, it was rather to understand why the 
problems appear hard for some local algorithms, in some intervals of 
ensemble parametrization. These efforts 
culminated in designing a new polynomial algorithm, {\it survey propagation} 
\cite{MPZ02, MZ02}, which empirically outspeeds all the previously known 
heuristics. Rigorous undertanding of this algorithm is, however, still 
missing.  

The fact that lies behind this success is the intrinsic similarity of
the combinatorial optimization problems to physical systems called
spin glasses \cite{MPkniha}. The organization of solutions is 
analogous to the structure of the free energy landspace of the
physical models.  Several phases can be located in the parameter
space, with abrupt transitions between the different phases.  An example is
the SAT/UNSAT transition, 
i.e.~the transition from the SATisfiable (SAT) phase where almost every 
instance is satisfiable (ground state of energy zero) to the 
UNSATisfiable (UNSAT) phase where almost
every instance is unsatisfiable (positive ground-state energy). Another is
the glassy transition where the phase space splits into many clusters
and metastable states, and where many dynamical procedures (a physical
dynamics or an algorithm) are unable to find the ground state. This
connection between the structure of solution and the average
algorithmic performance was the main motivation for detailed studies
of the phase diagram of the 
$K$-satisfiability \cite{BMW00, MPZ02, MZ02, MPR03}, 
Vertex Coloring \cite{MPWZ02, KPW04} and many other problems.

The presented study of the phase diagram of the 1-in-$K$
satisfiability (sometimes called \emph{exact satisfiability} \cite{RSK92}) 
problem
adds one more item to this list. But prolonging a list is not the 
main motivation for this work. 1-in-$K$ SAT on the symmetric ensemble 
(probability that a variable in a clause is negated is $1/2$) is one 
of the few NP-complete problem which has been proven to be on average 
polynomial (Easy) \cite{ACIM01}. On the other hand for the positive ensemble 
(no negations, equivalent to Exact Cover) no such proof exists, nor is there 
a heuristic algorithm with empirically polynomial 
time performance in the vinicity of the SAT/UNSAT transition.
However, by analogy with K-SAT and coloring, we may expect polynomial time performance in this region using survey propagation.

Our main motivation is to interpolate between the symmetric and
positive ensemble to show how the phase space changes. For this reason
we introduce a $\epsilon$--1-in-$K$ SAT problem, and study the phase
diagram in parameters $(\gamma,\epsilon)$, where $\epsilon$ stays for
the probability that a variable in a clause is negated, and $\gamma$
is the average connectivity of a variable.  To our knowledge, this
general ensemble is considered here for the first time.  We generalize
the rigorous probabilistic analysis to the general $\epsilon$
case. Then we use the replica-symmetric and
one-step--replica-symmetry-breaking cavity methods \cite{MP99,MP03} to
understand more features
of the problem in the whole space of parameters.

Our motivation is similar to the one which led to the introduction of the
$(2+p)$-SAT problem \cite{MZ98,MZKST99, Achlioptas}, where the instances are a mixture of 2-SAT and 3-SAT clauses on Erd\"os-R\'enyi graphs.
A parameter $p$ interpolates between the ensemble of random 3-SAT
formulas, which are
know to be computationally hard \cite{MPZ02,MZ02} in a region including 
the SAT/UNSAT
transition, and random 2-SAT formulas, for which an anycase polynomial
algorithm exists. A statistical physics approach has been applied 
to study the $(2+p)$-SAT problem, however only the replica-symmetric 
solution was investigated. Analogical interpolation between P and 
NP-complete cases of some other problems have been investigated 
in~\cite{2_p,2pcol}.

\subsection{The model}
\label{model}

A \emph{factor graph} $G=(V_v,V_c;E)$ is a bipartite graph, where the
two species of vertices are called respectively \emph{variables} $i
\in V_v$ and \emph{clauses} $a \in V_c$. It is a common graphical
object used in computer science in order to encode the geometrical
framework of a problem in combinatorial optimization, as it often
allows us to shorten and clarify the ``rules of the game''.

This is the case also for 1-in-$K$-SAT.  Indeed, cases of both $K$-SAT
and 1-in-$K$-SAT are example of boolean satisfiability problems, and
thus formally inscribed in a framework of boolean logic expressions:
we deal with $M$ boolean clauses over $N$ variables, which should be
simultaneously satisfied (i.e.~evalued to True), in order to consider
the $N$-tuple of assignments to the variables a \emph{solution} of the
problem instance. Each clause $a$ involves $K$ out of the $2N$
literals $x_1, \ldots, x_N, \ngg x_1, \ldots, \ngg x_N$ (not $x_i$ and 
$\ngg x_i$ simultaneously). While a $K$-SAT clause is satisfied if 
\emph{at least} one of the involved literals is True,
a 1-in-$K$-SAT clause is satisfied if \emph{exactly} one of the
involved literals is True. 

For both problems, a factor graph $G$ (whose clause-nodes $a$ have
degree $K$) and a function $J:E(G) \to \pm 1$ fully encode an
instance: if clause $a$ involves literal $x_i$ or $\ngg x_i$ we will
have an edge $(i,a) \in E(G)$, and $J_{ai}=+1$ or $-1$
respectively. It is customary to draw edges with $J=+1$ as solid
lines, and edges with $J=-1$ as dashed lines.

If we use the common identification with ``spin variables'' $s_i$
\[
\begin{array}{ccc}
x_i=\textrm{True}  & \qquad \longleftrightarrow \qquad & s_i=+1 \\
x_i=\textrm{False} & \qquad \longleftrightarrow \qquad & s_i=-1
\end{array}
\]
the function $E_{\{J_a\}}(s)$ corresponding to a 1-in-$K$-SAT clause is
\be
\label{eq.Ebool}
E_{\{J_a\}}(s)=
\left\{
\begin{array}{lcl}
\textrm{True}  && s_{i_1} J_{a i_1} + \cdots + s_{i_K} J_{a i_K} = 2-K \\
\rule{0pt}{12pt}%
\textrm{False} && \textrm{otherwise}
\end{array}
\right.
\ee
and we say that $s$ is a solution of the given instance
if $\bigwedge_a E_{\{J_a\}}(s) = \textrm{True}$.

1-in-$K$-SAT is polynomial in the case $K=2$
(coinciding with 2-XOR-SAT, or 2-Coloring), while it
is NP-complete for $K \geq 3$, even in the restriction to all $J$'s
positive (unlike $K$-SAT).

For what concerns average-case complexity,
two Erd\"os-R\'enyi-like random ensembles of instances are commonly 
considered. In both cases we have $N$ variables and every possible
clause is present with 
probability $p$ such that the average number of clauses is
$M=N \gamma/K$, and variables have Poissonian degree with average
$\gamma$.
Then we distinguish
\begin{description}
\item[Positive Poisson ensemble:] The edge parameters $J_{ai}$ are all $+1$. 
In this case we use a shorthand for the energy function $E_{J_a=(+,+,+)}=E_a$.
\item[Symmetric Poisson ensemble:] The edge parameters $J_{ai}$ are random 
  independent in $\{\pm 1 \}$ with equal probability.
\end{description}

\noindent
The positive version of the problem is the one which corresponds to
Exact Cover, in the case of incidence matrices whose columns have $K$
nonzero entries. 

In this paper we study the generalization to the ensemble in which the
$J$'s are taking value $\pm 1$ independently, $\epsilon \in [0,1/2]$
being the probability of having $J=-1$.  We call this
generalization $\epsilon$--1-in-$K$ SAT, in order not to confuse with
the 1-in-$K$ SAT by which is often meant only the symmetric ensemble.
We describe the phase diagram of this problem in the 
parameters $(\epsilon, \gamma)$.

\subsection{Main results and the paper organization}

Throughout this paper we present methods and results relevant to the
problem of $\epsilon$--1-in-$K$-SAT with $K=3$ only. Generalisation to
instances of larger clause length requires, in most cases, only small
changes in methodology.

In section \ref{bounds} and appendices \ref{app_UC}-\ref{other_bounds} 
we derive algorithmic and probabilistic bounds, both rigorous, for 
the SAT/UNSAT threshold in the $\epsilon$--1-in-3-SAT problem.
The most remarkable result of those sections is that the bound is 
tight for $\epsilon \in [0.2726,1/2]$, so in that interval 
the SAT/UNSAT threshold is rigorously known. This generalizes 
the result of \cite{ACIM01} for the symmetric ensemble $\epsilon=1/2$.

In section \ref{RS_cav} we develop the Replica-Symmetric (RS) solution.
First we write the replica-symmetric equations \ref{RSce}, then we discuss the 
zero-temperature limit \ref{cavityZT}. We analyze the hard-fields
solution in \ref{RS_hard}, and the soft-fields solution in
\ref{entropic_cav}. 
However, as we show in \ref{RS_stab}, this solution cannot be correct
(cease to be stable) above a certain connectivity not larger than the expected SAT/UNSAT 
transition. At this connectivity the belief propagation algorithm 
would fail to converge. In fact, there even exists a region in the phase 
diagram, where the RS solution is not stable, and yet the Short Clause
Heuristics (SCH) algorithm is proven to work in on average polynomial time. 
To our knowledge this does not happen in any of the previously studied 
models, and is a point worth further investigation. 

In section \ref{1RSB_cav} we work out the
one-step--Replica-Symmetry-Breaking (1RSB) solution.
In this case we assume the existence of many disconnected clusters of
solutions, and many metastable states, which can actually trap most of 
the traditional algorithms. 
We write the general equations in section \ref{G1RSBce}, then we concentrate 
on the zero-temperature zero-energy case \ref{ytoinfty}, which leads 
to the survey propagation equations. The zero-temperature positive-energy
case is studied in appendix \ref{yfinite}. In appendix \ref{S1RSB} 
we check the local stability of the 1RSB solution. 

\begin{figure}[!ht]
\caption{\label{fig_first} The phase diagram of $\epsilon$--1-in-3 SAT
  problem, for what concerns the SAT/UNSAT transition.  The parameters
  $\epsilon$ and $\gamma$ describe the probability of negations and
  the average variable connectivity.  For $\epsilon > 0.2726$, the
  threshold is rigorously $\gamma^*(\epsilon) = 1/\big(
  4\epsilon(1-\epsilon) \big)$ (drawn as a solid line), since the
  unit-clause upper bound 
  and
  short-clause-heuristic lower bound 
  coincide in that region.  For $\epsilon < 0.2726$, the dot-dashed,
  dashed and dotted line denore respectively the SCH lower bound, and
  the UC and first-moment-method (1MM) upper bounds. The solid line is
  our one-step replica-symmetry-breaking (1RSB) prediction for the
  SAT/UNSAT threshold. For $0 \le \epsilon<0.07$ the 1RSB result is
  stable (gray shading) and so the threshold is likely to be
  exact. For $0.07<\epsilon<0.2726$ the 1RSB result is unstable, and
  so the threshold is just approximate (expected to be an upper
  bound).  }
\begin{center}
\setlength{\unitlength}{50pt}
\begin{picture}(5,3.8)
\put(-0.1,-0.03){\includegraphics[scale=1, bb=0 0 270 250]
{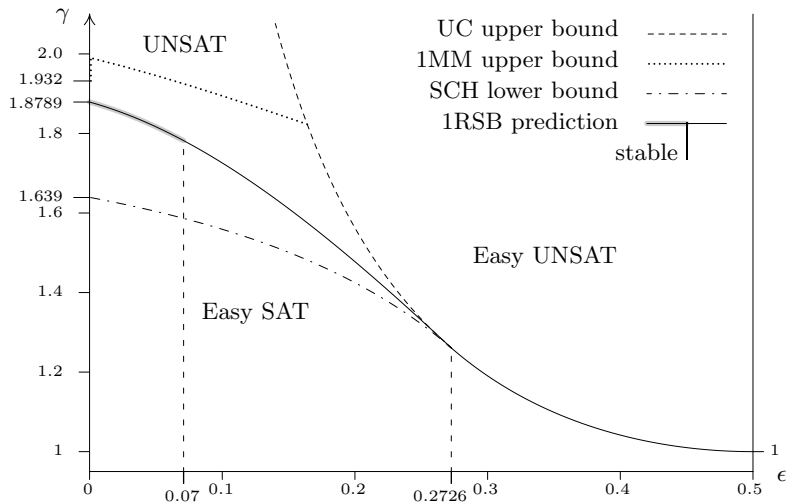}}
\put(0.05,0){${\scriptstyle 0}$}
\put(1,0){${\scriptstyle 0.1}$}
\put(2,0){${\scriptstyle 0.2}$}
\put(3,0){${\scriptstyle 0.3}$}
\put(4,0){${\scriptstyle 0.4}$}
\put(5,0){${\scriptstyle 0.5}$}
\put(2.55,-0.05){${\scriptstyle 0.2726}$}
\put(0.67,-0.05){${\scriptstyle 0.07}$}
\put(5.3,0.1){$\epsilon$}
\put(-0.05,3.6){\makebox[0pt][r]{$\gamma$}}
\put(-0.1,0.3){\makebox[0pt][r]{${\scriptstyle 1 }$}}
\put(-0.1,0.9){\makebox[0pt][r]{${\scriptstyle 1.2}$}}
\put(-0.1,1.5){\makebox[0pt][r]{${\scriptstyle 1.4}$}}
\put(-0.1,2.1){\makebox[0pt][r]{${\scriptstyle 1.6}$}}
\put(-0.1,2.7){\makebox[0pt][r]{${\scriptstyle 1.8}$}}
\put(-0.1,3.3){\makebox[0pt][r]{${\scriptstyle 2.0}$}}
\put(5.25,0.3){\makebox[0pt][l]{${\scriptstyle 1 }$}}

\put(-0.1,2.217){\makebox[0pt][r]{${\scriptstyle 1.639}$}}
\put(-0.1,2.925){\makebox[0pt][r]{${\scriptstyle 1.8789}$}}
\put(-0.1,3.105){\makebox[0pt][r]{${\scriptstyle 1.932}$}}

\put(4.1,3.466){\makebox[0pt][r]{UC~upper bound}}
\put(4.1,3.233){\makebox[0pt][r]
{1MM~upper bound}}
\put(4.1,3.0){\makebox[0pt][r]{SCH~lower bound}}
\put(4.1,2.766){\makebox[0pt][r]{1RSB prediction}}
\put(4.62,2.533){\makebox[0pt][r]{stable\ \rule{0.25pt}{14pt}}}

\put(3.0,1.75){Easy UNSAT}
\put(0.95,1.33){Easy SAT}
\put(0.5,3.35){UNSAT}
\end{picture}
\end{center}

\end{figure}

The main result of the 1RSB analysis is the prediction for the SAT/UNSAT
threshold, fig. \ref{fig_first}. For $\epsilon<0.07$ the 1RSB approach 
is stable around the SAT/UNSAT line, so the threshold is likely to be exact. 
Whereas for  $0.07<\epsilon<0.2726$ the 1RSB result is unstable, and a more 
involved analysis would be required to locate exactly the SAT/UNSAT threshold 
(the 1RSB result is expected to be an upper bound).
The presence of a nontrivial 1RSB solution in the small-$\epsilon$ 
region suggests the presence of a Hard-SAT region. Details of these 
results are discussed in section \ref{1RSBres}. 

\section{Rigorous bounds on the SAT/UNSAT threshold}
\label{bounds}

\subsection{Unit-Clause Propagation Analysis}

Unit-Clause (UC) algorithms are a class of randomized algorithms for
boolean satisfiability problems, which when applied to a specific instance,
seek a solution or a
certificate of unsatisfiability by assigning
variables to $\pm 1$ (or ``True/False'') whilst maximizing the amount of
logical deductions coming from uniquely determined constraints
(\emph{unit clauses}). In the absense of immediate deductions, variables
are fixed by some heuristic rule, and these free steps determine
a branching process on the space of feasible configurations.
Our analysis of unit-clause propagation is elucidated
in appendix \ref{app_UC}, while for a more general
description consider~\cite{Deroulers:CUU}.

Algorithms based on unit-clause propagation have been already analysed
for the problems of symmetric and positive 1-in-$K$ SAT.
For the positive ensemble ($\epsilon=0$) the best known
lower bound to the SAT/UNSAT transition is $\gamma=1.638$~\cite{MK05},
and no upper bound is known from unit-clause algorithms. For the
symmetric ensemble ($\epsilon=1/2$), the method allows us to determine that 
the exact SAT/UNSAT transition is $\gamma=1$~\cite{ACIM01}.  Here
we extend these results to compute the upper bound 
$\gamma_{\rm uc}(\epsilon)$ and the lower bound 
$\gamma_{\rm sch}(\epsilon)$ for
general probability of negation, which describe regions that are
almost surely (a.s.) Easy SAT or Easy UNSAT for an instance of the
$\epsilon$--1-in-3-SAT, sampled from the random $(\epsilon,\gamma)$
ensemble (cfr.~section \ref{model}).

In appendix \ref{ssec.UB} we demonstrate the upper bound 
$\gamma_{\rm uc}(\epsilon)$  to the connectivity above which an Easy UNSAT phase
exists.  Whenever $\gamma > \gamma_{\rm uc}(\epsilon)$, the instance is
a.s.~proven to be UNSAT by a randomized linear-time decimation algorithm
in which one tests, for all variables $i$, if both fixing $s_i=+1$ or
$s_i=-1$ lead to contradictions through unit-clause implications
alone. This line has the analytic form
$\gamma_{\rm uc}(\epsilon)=1/\big( 4 \epsilon (1-\epsilon) \big)$.

In appendix \ref{ssec.LB} we obtain the lower bound
$\gamma_{\rm sch}(\epsilon)$ to the connectivity below which an Easy SAT phase
exists. Now, we perform an extensive number of free choices, and thus
we should specify our
heuristic rule. It turns out that, among those tested, the 
Short Clause Heuristics
(SCH; assigning a variable in one of the shortest clauses remaining) is
the one attaing the best bound on the whole interval of $\epsilon$.
If $\gamma < \gamma_{\rm sch}(\epsilon)$, by fixing variables
according to SCH one can find a solution with finite probability on any
run. Restarting the procedure many times allows us to find a solution in
on average linear time. This extends the idea employed for Exact Cover
in \cite{MK05}. At all $\epsilon$ we find lower bounds by numerical
integration (see above figure), including 
$\gamma_{\rm sch}(0)=1.639$, consistent with the analysis of~\cite{MK05}.

Finally in appendix \ref{ssec.ESU} we prove analytically that on
the interval $\epsilon \in [0.2726, 1/2]$ the curves
$\gamma_{\rm uc}(\epsilon)$ and $\gamma_{\rm sch}(\epsilon)$ coincide.
The result includes the symmetric ensemble, for which it was originally
proven in \cite{ACIM01}. 
This fact indicates that there exists a region of the phase
diagram in which typical instances of the $(\epsilon,\gamma)$ ensemble
are easily solved, except at the exactly determined SAT/UNSAT transition
line.

\subsection{Upper bounds for small $\epsilon$}

For $\epsilon \to 0$ we have $\gamma_{\rm uc} \sim \epsilon^{-1}$ and
so we would like to find a
better upper bound by some different method.  An improvement is obtained
through the First Moment Method (1MM) on the 2-core of the
graph. Restriction to the 2-core makes the bound tighter, as it
reduces instance-to-instance fluctuations. This provides a line
$\gamma_{\rm 1mm}(\epsilon)$, 
which is finite
everywhere, and thus beats $\gamma_{\rm uc}$ in some interval
of small $\epsilon$ (details are in appendix \ref{app_1st}).
The best known upper bound for the positive ensemble ($\epsilon=0$) 
is $\gamma=1.932$, obtained by a refinement of the first moment 
method \cite{KSM04}.

Still, the first moment method is only probabilistic, and does not
allow us to find a certificate in polynomial time for a given instance. Such a
task is achieved at finite $\gamma$, also in the region of small
$\epsilon$ and \hbox{$\epsilon=0$}, 
through the embedding of 1-in-3-SAT into an instance of
3-XOR-SAT. While $E_{a}^{\textrm{1-in-3}}(s_1,s_2,s_3) = \textrm{True}$
on the three configurations $(s_1,s_2,s_3) = (+,-,-)$, $(-,+,-)$ and
$(-,-,+)$, the function $E_a^{\textrm{3-XOR}}(s_1,s_2,s_3)$ also
allows for $(s_1,s_2,s_3) = (+,+,+)$. So all the constraints are
linear relations, and the problem is formally solved by Gaussian
elimination. 
This gives an upper bound for an ``Easy-UNSAT'' phase, independently
of $\epsilon$ at $\gamma = 3 \alpha^* = 2.754$ where $\alpha^*$ is the
SAT/UNSAT threshold (clause-to-variable ratio) in 
3-XOR-SAT \cite{fede3xor,MRZ03}  
(cfr.~appendix \ref{XOR-SAT} for details).
Note for comparison that in random 3-SAT at given finite $\alpha$ (however large) in
the UNSAT phase, there is no polynomial algorithm
which can find a.s.~a certificate for a typical instance, and intuition 
strongly suggests that such an algorithm can not exist~\cite{CSz}.

\section{RS cavity approach}
\label{RS_cav}

The cavity method is developed within a statistical mechanics formulation. For this purpose we choose an integer-valued ``energy function'' for a single clause, 
$E_{\{J_a\}}(s)$, to be associated to the original boolean-valued
function $E_{\{J_a\}}^{\textrm{bool}}(s)$ in 
(\ref{eq.Ebool}); just as
$E_{\{J_a\}}(s)=0$ or $1$ respectively if
$E_{\{J_a\}}^{\textrm{bool}}(s) = \textrm{True}$ or
$\textrm{False}$.
We thus have a Hamiltonian
\be
\ham (s) 
= \sum_a 2 E_{\{J_a\}}(s)
\ef,
\label{eq.ham}
\ee
which counts (twice) the number of contradictions.  The factor
$2$ is a useful convention, so that all the cavity parameters to be 
introduced below (\emph{cavity fields} and \emph{biases}) will be 
integer in the ``zero-temperature''
limit.

The introduction of the cost function above allows us to define a
Gibbs weight $e^{-\beta \ham(s)}$, where $\beta$ is some parameter
(\emph{inverse temperature}), so that a single contradiction causes a dump of a
factor $e^{-2 \beta}$ in the measure of the configuration.  As
customary in statistical mechanics, one introduces a partition
function
\be
Z(\beta)=\sum_s e^{-\beta \ham(s)}
\ee
and a set of observables (say, probabilities of having patterns $A$)
\be
\mathrm{prob}(A) := Z^{-1} 
\!\!\!\! \sum_{s: \; A \textrm{ happens}} \!\!\!\!
e^{-\beta \ham(s)}
\ef.
\ee

Within this framework the cavity method translates 
certain obvious recurrence relations for
interaction structures on a factorized graph (tree)
to approximate self-consistent
equations for local expectation values on a graph which is
only ``locally tree-like'' (e.g.~a sparse Erd\"os-R\'enyi graph at large 
$N$, where loops are expected to arise at lengths of order $\ln N$).

The replica-symmetric (RS) assumption is used at a certain point. It
consists in assuming that there is a single pure state describing the
equilibrium behaviour of the ensemble. In turns, it will allow to
neglect certain connected correlation functions.  In this section we
develop the cavity method under this hypothesis, while extensions are
discussed in later sections.

We will not review in detail all the derivations of the equations, 
instead we just introduce, in section \ref{RSce}, some notations on
the ``easy'' case of interaction on a tree, and give without proof the
further formulas which are valid in the various contexts. A heuristic
consideration of the complications arising on a random graph with long
loops can be found in~\cite{MP99, MP03}.

\subsection{RS cavity equations}
\label{RSce}

\begin{figure}[!ht]
\setlength{\unitlength}{50pt}
\begin{picture}(4.4,2.6)
\put(0,0){\includegraphics[scale=1, bb=0 105 220 235, clip=true]
{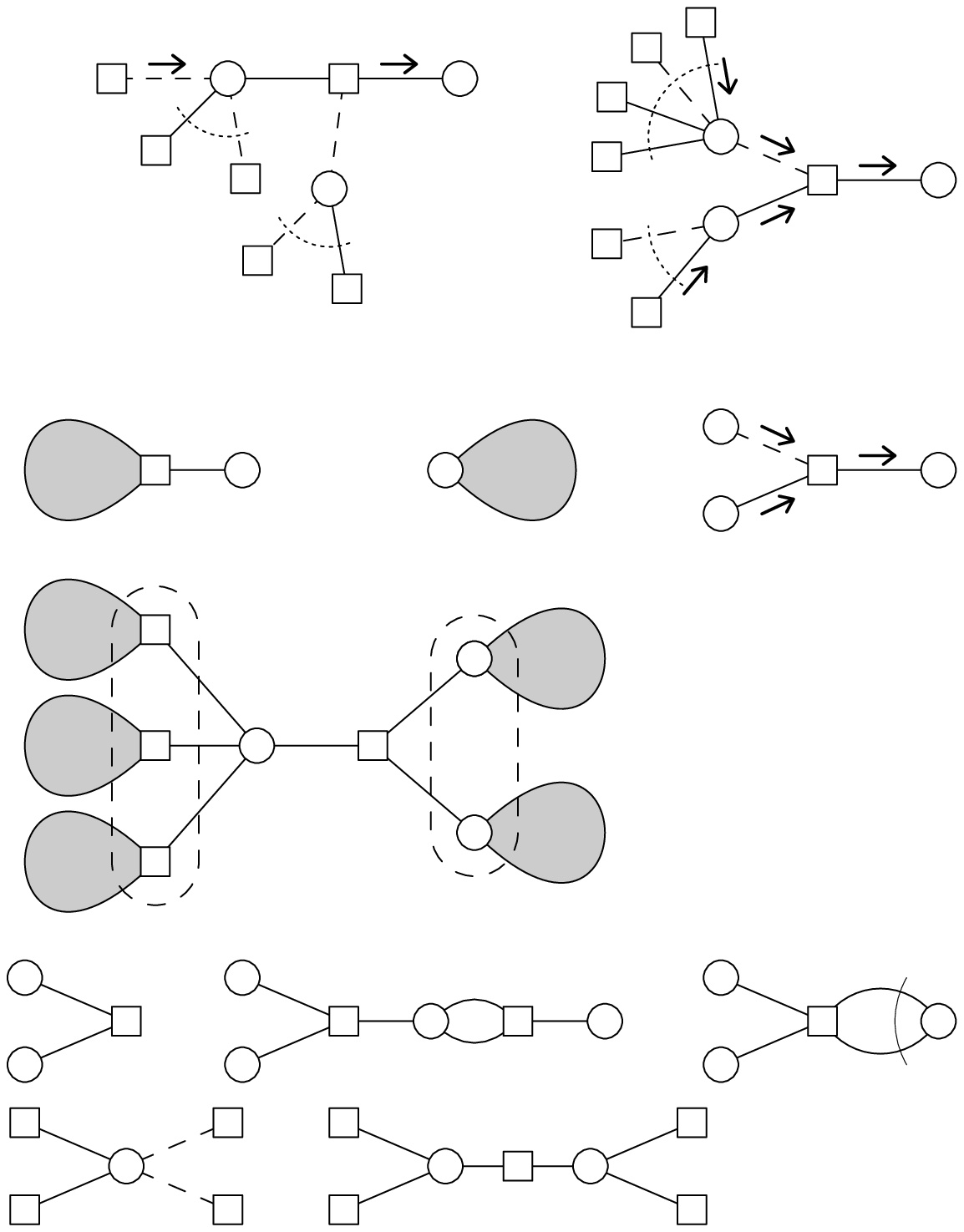}}
\put(1.05,1.8){$b$}
\put(1.75,1.5){$i$}
\put(2.55,1.5){$a$}
\put(3.25,1.6){$j$}
\put(3.25,0.9){$k$}
\put(3,2.35){$\partial a \setminus i$}
\put(0.8,2.55){$\partial i \setminus a$}
\end{picture}
\quad
\setlength{\unitlength}{40pt}
\begin{picture}(4.8,1)(-1.5,0)
\put(-0.5,0.5){$Y_{s_i}^{b \to i} :$}
\put(0,0){\includegraphics[scale=1, bb=0 240 100 290, clip=true]
{fig_smallFG.eps}}
\put(1.35,0.9){$b$}
\put(2.1,0.9){$i$}
\put(1.5,0.05){fixed to $s_i$}
\put(1.9,0.22){\vector(1,2){0.14}}
\put(-0.5,2.4){$Z_{s_j}^{j \to a} :$}
\put(0.1,1.9){\includegraphics[scale=1, bb=140 240 210 290, clip=true]
{fig_smallFG.eps}}
\put(0.45,2.8){$j$}
\put(0.,1.95){fixed to $s_j$}
\put(0.3,2.07){\vector(1,2){0.14}}
\end{picture}
\caption{\label{newfig} Left: Factor graph representing the 1-in-3 SAT
  problem, in a neighbourhood of an edge $(i,a)$.
Right: Definition of the cavity partition functions.}
\end{figure}

Consider a problem defined on a factor graph $G$, such that, for a
certain edge $(i,a)$, $G$ is composed of factorized components
attached to the vertices in a neighbourhood of radius 1 of $(i,a)$.
Call $(\partial i \setminus a)$ the set of other clauses, besides $a$,
neighbouring $i$, and $(\partial a \setminus i)$ the set of other
variables, besides $i$, neighboring $a$. This description motivates a
factor graph of the form on fig. \ref{newfig} left, where ``gray bubbles'' stands 
for some other parts of the graph, and there are no paths connecting 
distinct bubbles, except through the explicitly drawn neighborhood of $(i,a)$.


For a variable $j \in (\partial a \setminus i)$, call $Z_{s}^{j\to a}$
the quantity $Z \cdot \textrm{prob}(s_j=s)$ on the system consisting of the
(gray-bubble) subgraph attached to vertex $j$, see fig. \ref{newfig} right, 
$Z$ being the partition function of this subsystem. Similarly for a clause 
$b \in (\partial i \setminus a)$, 
call $Y_{s}^{b \to i}$ the quantity $Z \cdot \textrm{prob}(s_i=s)$
on the system consisting on the subgraph attached to node $i$ through
edge $(b,i)$, see fig. \ref{newfig} right.
%
Then we have the composition relations (say 
$\partial a \setminus i = \{j,k\}$ for our clause of degree 3)
\begin{subequations}
\label{part_sum}
\begin{align}
   Z_{s_i}^{i\to a} &= \prod_{b\in \partial i \setminus a}  
                          Y_{s_i}^{b\to i} \, ,
\\
   Y_{s_i}^{a\to i} &= \sum_{s_j,s_k} e^{-2 \beta E_{\{J_a\}}(s_i,s_j,s_k)}
                          Z_{s_j}^{j\to a} Z_{s_k}^{k\to a} \, .
\end{align}
    \label{eq.BP_Z} 
\end{subequations}
At this stage we note that, in rewriting the cavity partition sums in
terms of probabilities, the \emph{belief propagation} equations
\cite{pearl,YFW03}, familiar to computer scientists, are attained.

As usual in physics we reparametrize the pairs $Z_{+}, Z_{-}$ as 
different natural quantities, a \emph{free energy} 
$F=1/\beta \; \ln (Z_+ + Z_-)$ and a
\emph{magnetic field} $h=1/(2\beta) \; \ln (Z_+ / Z_-)$, the field
that, if applied to the variable in substitution of the whole system, would
cause the same average magnetization.
We define the \emph{cavity fields} and \emph{cavity biases}
in the following way
\begin{align}
  e^{2\beta h_{i\to a}} & := \frac{Z_{+}^{i\to a}}{Z_{-}^{i\to a}} \, ,
&
  e^{2\beta u_{a\to i}} & := \frac{Y_{+}^{a\to i}}{Y_{-}^{a\to i}} \, .
\label{def_hu}
\end{align}
The recursion equations for $h$ and $u$ then follows from (\ref{part_sum})
\begin{subequations}
\label{BP_T}
\begin{align}
  h_{i\to a} & = \sum_{b\in \partial i \setminus a} u_{b\to i} \, , \\ 
  u_{a\to i} & = \frac{1}{2 \beta} \ln 
       \frac{\sum_{s_j,s_k} \exp \big[ \beta ( 
      h_{j\to a} s_j + h_{k\to a} s_k - 2 E_{\{J_a\}}(+1,s_j,s_k) ) \big] }
          {\sum_{s_j,s_k} \exp \big[ \beta (
      h_{j\to a} s_j + h_{k\to a} s_k - 2 E_{\{J_a\}}(-1,s_j,s_k) ) \big] }\, .
\end{align}
\end{subequations}
One can think of $u$'s and $h$'s as messages being attached to the
edges of the graph, and oriented (the $u$'s towards the variable, the
$h$'s towards the clause). Then, the update functions (\ref{BP_T})
are represented on the variable- and clause-nodes, as in the figure
below.
\[
\setlength{\unitlength}{50pt}
\begin{picture}(2,1)
\put(0.1,0){\includegraphics[scale=1, bb=240 240 335 290, clip=true]
{fig_smallFG.eps}}
\put(2,0.45){$i$}
\put(1.15,0.35){$a$}
\put(0,0.85){$j$}
\put(0,0.1){$k$}
\put(0.5,0.){$h_{k\to a}$}
\put(0.5,0.95){$h_{j\to a}$}
\put(1.18,0.78){$u_{a\to i}$}
\end{picture}
\]
Similarly, we handle the free energies.
First define the accessory quantities
\begin{subequations}
\label{eq.freeen}
\begin{align}
   Z^{i} &= \prod_{a\in \partial i}  Y_{+}^{a\to i} 
            + \prod_{a\in \partial i}  Y_{-}^{a\to i} \, ,\\ 
   Y^{a} &= \sum_{s_i,s_j,s_k} e^{-2 \beta E_{\{J_a\}}(s_i,s_j,s_k)}
            \; Z_{s_i}^{i\to a} Z_{s_j}^{j\to a} Z_{s_k}^{k\to a} \, .
\end{align}
\end{subequations}
Then we write the free-energy shift $\Delta F^{a \cup \da}$ after
adding a clause $a$ and connecting all its neighbors $i \in \da$, 
and free-energy shift $\Delta F^{i}$ after connecting all the
components incident on variable $i$
\begin{subequations}
\label{free_s}
\begin{align}
\label{free_s1}
\begin{split}
e^{-\beta \Delta F^{a \cup \da} } 
&=
\frac{Y^{a}}{\prod_{i\in \da} \prod_{b\in \di \setminus a}
    ( Y_{+}^{b\to i} +  Y_{-}^{b\to i}) } \, , \\
&= \frac{ \sum_{s_i,s_j,s_k}
    \exp \big\{ \beta \big[ h_{i\to a}s_i+ h_{j\to a}s_j + h_{k\to a}s_k 
   - 2 E_{\{J_a\}}(s_i,s_j,s_k) \big] \big\} }
   { \prod_{i\in \da} \prod_{b\in \di \setminus a}
    2 \cosh (\beta u_{b\to i} )} \, ,
\end{split}
\\
e^{-\beta \Delta F^{i} }
&=
\frac{Z^{i}}
  {\prod_{a\in \partial i}( Y_{+}^{a\to i} + Y_{-}^{a\to i})} 
=
\frac{2 \cosh ( \beta \sum_{a\in \partial i} u_{a\to i} ) }
{\prod_{a\in \partial i} 2 \cosh ( \beta u_{a\to i} )} 
\ef.
\label{free_s2}
\end{align}
\end{subequations}
Finally, for the free energy, $F = 1/\beta \ln
Z(G)$, of a tree graph one gets
\begin{equation}
 F(\beta) = \sum_a \Delta F^{a \cup \da} - \sum_i (d_i-1) \Delta F^{i}
\ef,
    \label{free}
\end{equation}
where $d_i$ is degree of variable $i$. Writing this in terms of fields
we note the cancellation of the factors $2 \cosh ( \beta u_{a\to i} )$
between the denominators of (\ref{free_s1}) and (\ref{free_s2}).
Furthermore, in the numerator of (\ref{free_s2}), the combination
$h_i := \sum_{a\in \partial i} u_{a\to i}$ appears.
This is the ``total field'' parameter for the
magnetization of variable $i$ in cavity approximation.

The free energy as a function of
inverse temperature $\beta$ on a given graph
allows us to determine, by Legendre transform, the number
$\exp\big(S(E)\big)$ of configurations of given energy
$E$ (number of violated clauses). Both $S$ and $E$ are
extensive, i.e.~of order $N$, and corrections decreasing with $1/N$
are understood.
\begin{align}
  E(\beta) & = \frac{\partial (\beta F(\beta))}{\partial \beta} 
\ef;
&
  S(E) & = (E-F) \; \beta(E) 
\ef.
 \label{energy_entropy}
\end{align}
The main insight here is that we can think of the set of cavity fields
as parameterization for the local ``magnetizations'' of variables $i$
(i.e.~probability of being
$s_i=+1$, for two-state variables), in a system in which the
interaction of $i$ with a neighboring clause $a$ has been modified
(\emph{cavity system}). If the clause $a$ 
has been ``switched off'', the nodes
in the neighborhood of $a$ now become well-separated on the graph
which effectively describes the cost function. An assumption of
decorrelation of variables, which is exactly true for variables in
disconnected components, and approximatively valid for variables
sufficiently far away on the graph, out of a critical temperature and
\emph{within a pure thermodynamic phase}, provides us
self-consistent equations for the cavity fields.
The equations are exactly the same as those 
we wrote for factorized graphs and hold in the leading order in the 
system size $N$, in particular eq. (\ref{free}), see~\cite{MP99}.  
The cavity assumption
can be self-consistently checked as we will describe in 
section~\ref{RS_stab}.

\subsection{The zero-temperature limit: Hard and soft fields}
\label{cavityZT}

In the limit of zero temperature, $\beta \to \infty$, the update of
cavity biases (\ref{BP_T}) simplifies significantly to
\begin{subequations}
\label{eqs.cavityZT}
\begin{align}
  h_{i \to a} &= \sum_{b \in \di \setminus a} u_{b \to i} \, ,
\\
\begin{split}
  u_{a \to i} &= \frac{1}{2} \Big[
 \max_{s_j,s_k}
   \big( h_{j\to a} s_j + h_{k\to a} s_k -2 E_{\{J_a\}}(+1,s_j,s_k) \big) \\
 & \hspace{1.cm}
 - \max_{s_j,s_k}
  \big( h_{j\to a} s_j + h_{k\to a} s_k -2 E_{\{J_a\}}(-1,s_j,s_k) \big)
\Big]
\ef.
\end{split}
\label{BP_0}
\end{align}
\end{subequations}
It is immediately seen that, as $E_{\{J_a\}}(s)$ is evaluated over
$\{0,1\}$, it
is self-consistent to assume that $h \in \mathbb{Z}$ and 
$u \in \{-1,0,+1\}$.

In fact the only characteristic property of $E_{\{J_a\}}(s)$ we need to have 
is that $E_{\{J_a\}}(s)=0$ if and only if $s$ satisfies
clause $a$, and that $E_{\{J_a\}}(s_1, s_2, s_3) - E_{\{J_a\}}(-s_1,
s_2, s_3) \in \{-1,0,+1\}$, (a kind of discrete ``Lipschitz''
condition), which clearly holds for our choice of
Hamiltonian~(\ref{eq.ham}).

The only other choice of Hamiltonian for 
1-in-3-SAT sharing this property is
\be
\label{eq.ham'}
{\cal H}'(s)=2 \sum_a E'_{\{J_a\}}(s)
\ef,
\ee
where $E'_{a}(s)$ coincides with $E_{a}(s)$ except that on $(+,+,+)$,
where it is valued $E'=2$ instead of $1$ (because 2 flips are required
in order to satisfy the clause).

The fact that $h, u \in \mathbb{Z}$ is much more than
self-consistent, it is necessary, even for the ``true'' cavity fields
(the ones that we would find from the evaluation of global partition
functions, instead of the ones being solution of the cavity equations),
and approximatively true in a whole region of large $\beta$ (it
suffices that $\beta \gg \ln N$). Let us 
concentrate first on $h_{i \to a}$ and say that
$Z_+^{i \to a} = \sum_n g_+(n) e^{-2 \beta n}$, where the integer
coefficients $g_+(n)$ count the configurations with $n$ violated
clauses, in the proper cavity system labeled by $(i \to a)$. There
will be a certain value $n_+$ corresponding to the first non-vanishing
coefficient $g_+(n)$. Identical definitions are assumed for 
$+ \leftrightarrow -$. Then we have
\be
\label{eq.lowTfields}
h_{i \to a} 
=
\frac{1}{2 \beta} 
\ln \frac{ Z_+^{i \to a} }{ Z_-^{i \to a} }
= 
(n_- - n_+) + \frac{1}{2 \beta} \ln \frac{g_+(n_+)}{g_-(n_-)} + 
    \mathcal{O}(e^{-2 \beta}) \, .
\ee
So, at all orders in a purely algebraic expansion in powers of
$1/\beta$, we only have two terms: a first one, $(n_- - n_+)$, the
\emph{hard field}, is constrained to be integer; and secondly the
coefficient
in the second term, the \emph{soft field},
which being the logarithm of the ratio of two
(potentially large in $N$) integers, is simply taken as a value over
$\mathbb{R}$.

In particular the ground-state energy, $\beta\to \infty$ limit of 
eq. (\ref{energy_entropy}), can be computed using only the hard fields.
On the other hand, to compute the 
ground-state entropy the soft fields are necessary 
(the general relation follows from (\ref{energy_entropy}), but is 
rather lengthy). 

We will see in the next section that working only with the hard 
fields has huge computational advantages, however, as they do not 
contain all the information, we come back to the soft fields in
section \ref{entropic_cav}.

\subsection{The hard-fields analysis: Warning Propagation}
\label{RS_hard}

In this section we return to equations (\ref{eqs.cavityZT}), and
neglect for this moment the $1/\beta$ part of the field.  The
corresponding equations are called \emph{warning propagation}, and the
discrete set of possible values for the biases takes an interpretation
in terms of ``kinds of warnings'':

\begin{center}
\begin{tabular}{rc|cp{10cm}}
 $u_{a\to i}=-1$ &&&
  Clause $a$ tells to variable $i$: ``I think you should be $-1$''
\\
 $u_{a\to i}=0$  &&&
  Clause $a$ tells to variable $i$: ``I can deal with any value you
 take''
\rule{0pt}{12pt}
\\
 $u_{a\to i}=+1$ &&&
  Clause $a$ tells to variable $i$: ``I think you should be $+1$''
\rule{0pt}{12pt}
\end{tabular}
\end{center}

\noindent
The analogous interpretation for fields $h$ is

\begin{center}
\begin{tabular}{rc|cp{10cm}}
 $h_{i\to a}<0$ &&&
  Variable $i$ tells to clause $a$: ``I would prefer to be $-1$''
\\
 $h_{i\to a}=0$  &&&
  Variable $i$ tells to clause $a$: ``I don't have any strong preferences''
\rule{0pt}{12pt}
\\
 $h_{i\to a}>0$ &&&
  Variable $i$ tells to clause $a$: ``I would prefer to be $+1$''
\rule{0pt}{12pt}
\end{tabular}
\end{center}

\noindent
From which the prescriptions (\ref{eqs.cavityZT}) on how to
update the ``warnings'' over the graph also become intuitive.

We now determine statistics over the ensemble of random pairs $(G,\{J_a\})$. It
turns out that, although fields
$h$ can take infinite values, by virtue of the Poisson ensemble
the equations are closed under a finite
number of parameters:
We define probabilities $p_-$/$p_+$/$p_0$ that cavity fields $h$ are 
negative/positive/zero,
and similar probabilities $q_-$/$q_+$/$q_0$ that biases $u$
are $-1$/$+1$/$0$.
Then, the statistical average of equations (\ref{eqs.cavityZT}), seen
as defining a dynamics of time evolution for the distributions of the
fields, gives the following dynamical map over ${\mathbf{q}}=( q_+, q_- )$
(the auxiliary vector ${\mathbf{p}}=( p_+, p_- )$ is also defined)
\begin{subequations}
\label{eq.WP1}
\begin{align}
\label{eq.WP1a}
{\mathbf{q}}  &= M {\mathbf{q}'}
\ef;
&
{\mathbf{q}'} &= ( p_-^2, 2 p_+-2 p_+^2 )
\ef;
\\
\label{eq.WP1b}
{\mathbf{p}}  &= M {\mathbf{p}'}
\ef;
&
{\mathbf{p}'} &= (f(\gamma q_+,\gamma q_-),f(\gamma q_-,\gamma q_+))
\ef;
\end{align}
\end{subequations}
where we recall that $\gamma$ is the mean degree of variable, 
while matrix $M$ describes the probability of negations
\be
M=
\begin{pmatrix}
     1-\epsilon & \epsilon   \\
       \epsilon & 1-\epsilon 
\ef.
\end{pmatrix}
\ee
Finally the function $f(r,s)$ gives the probability that the
difference of two Poissonian-distributed integers (resp.~with rate
$r$ and $s$) is positive
\begin{align}
f(r,s) &:=
\sum_{m=0}^{\infty}
\sum_{n=m+1}^{\infty}
{\rm Poiss}_r(n) {\rm Poiss}_s(m) 
\ef;
&
{\rm Poiss}_{\alpha}(n) &:= e^{-\alpha} \frac{\alpha^n}{n!}
\ef.
\end{align}
The ``paramagnetic'' state $\mathbf{q} = \mathbf{0}$ is everywhere a
solution of (\ref{eq.WP1}). It is however numerically unstable above
the line $\gamma_{\rm uc}=1/\big( 4\epsilon(1-\epsilon) \big)$ (coinciding with
the unit-clause upper bound).  A non-paramagnetic 
$\mathbf{q} \neq \mathbf{0}$ solution appears continuously above this
line and is stable.  Conversely, for $\epsilon < \epsilon^*$, the
non-paramagnetic solution appears discontinuously at connectivity
$\gamma^*_{\rm RS}$, and is a stable local attractor.

\begin{figure}
\caption{\label{fig_RS}
 Results of the replica-symmetric cavity analysis and its stability. 
 The continuous curves with no bullets are the rigorous bounds
 left for comparison. 
 The line with diamonds corresponds to the replica-symmetric prediction for 
 the SAT/UNSAT threshold derived from the Hamiltonian ${\cal H}$, with
 $E_a(+,+,+)=1$, while the line with triangles to the one derived
 from the Hamiltonian ${\cal H}'$, with $E'_a(+,+,+)=2$.  
 The dotted line is the soft-fields instability of the 
 replica-symmetric solution, it is separating the stable region (below) from
 the unstable one (above). For $\epsilon>0.33\pm 0.02$ the stability 
 line seems to coincide with the SAT/UNSAT threshold.}
\begin{center}
\setlength{\unitlength}{30pt}
\begin{picture}(9.5,6.4)(0,-1.2)
\put(-0.1,-1.33){\includegraphics[scale=0.6, bb=0 0 270 250]
   {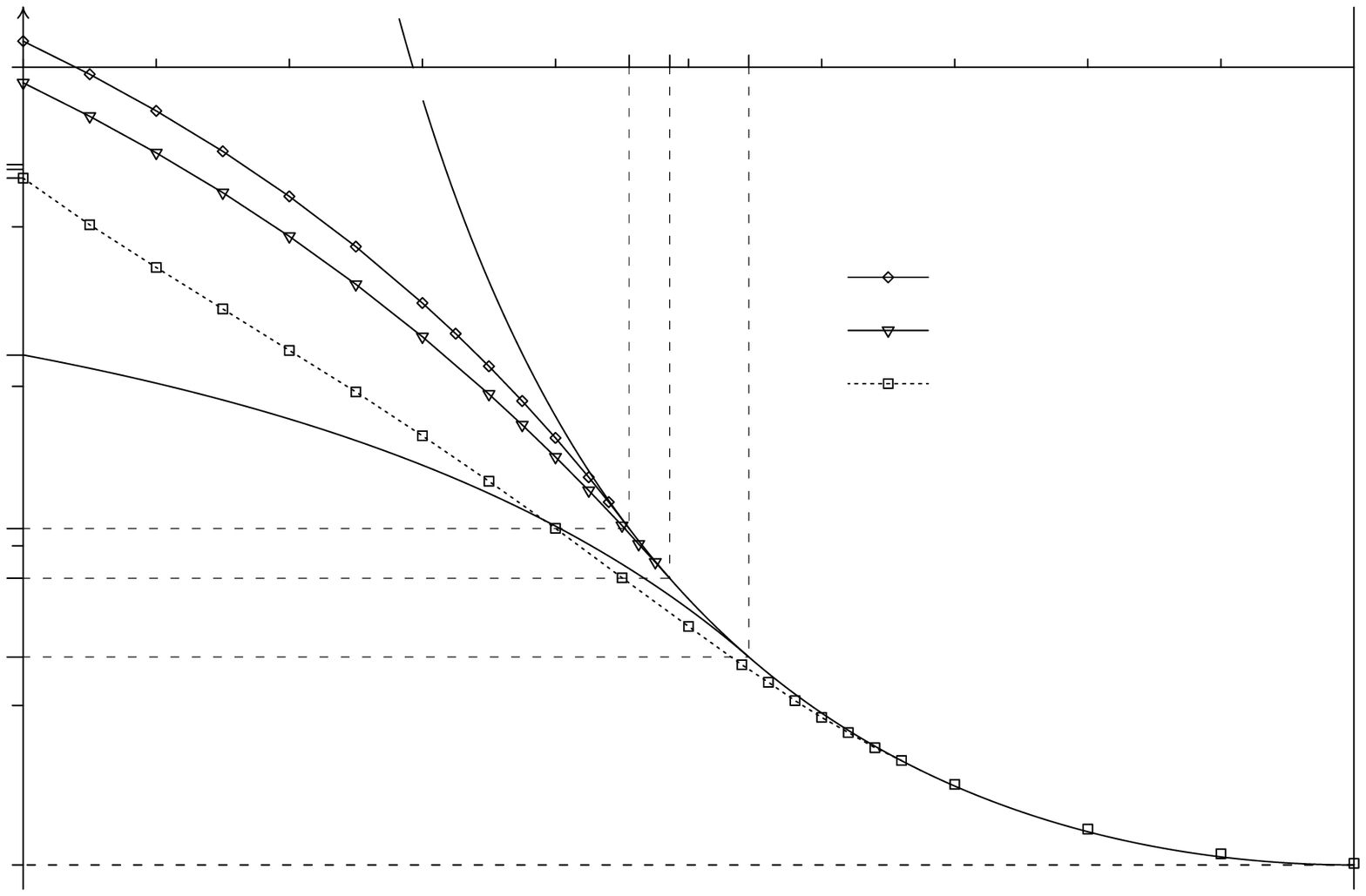}}
\put(1,4.3){${\scriptstyle 0.05}$}
\put(2,4.3){${\scriptstyle 0.1}$}
\put(3,4.3){${\scriptstyle 0.15}$}
\put(4,4.3){${\scriptstyle 0.2}$}
\put(5,4.3){${\scriptstyle 0.25}$}
\put(6,4.3){${\scriptstyle 0.3}$}
\put(7,4.3){${\scriptstyle 0.35}$}
\put(8,4.3){${\scriptstyle 0.4}$}
\put(9,4.3){${\scriptstyle 0.45}$}
\put(5.2,4.65){${\scriptstyle 0.2726}$}
\put(4.3,4.65){${\scriptstyle 0.2277}$}
\put(4.6,4.85){${\scriptstyle 0.2429}$}
\put(9.8,4.6){$\epsilon$}
\put(-0.,4.8){\makebox[0pt][r]{$\gamma$}}
\put(-0.1,-1.6){\makebox[0pt][r]{${\scriptstyle 1  }$}}
\put(-0.1,-0.4){\makebox[0pt][r]{${\scriptstyle 1.2}$}}
\put(-0.1, 0.8){\makebox[0pt][r]{${\scriptstyle 1.4}$}}
\put(-0.1, 2.0){\makebox[0pt][r]{${\scriptstyle 1.6}$}}
\put(-0.1, 3.2){\makebox[0pt][r]{${\scriptstyle 1.8}$}}
\put(-0.1, 4.4){\makebox[0pt][r]{${\scriptstyle 2.0}$}}

\put(-0.1,2.246){\makebox[0pt][r]{${\scriptstyle 1.639}$}}
\put(-0.1,3.72){\makebox[0pt][r]{${\scriptstyle 1.875\,(3)}$}}
\put(-0.1,3.5){\makebox[0pt][r]{${\scriptstyle 1.861}$}}



\put(7,3.6){Easy UNSAT}
\put(1.5,-0.8){Easy SAT}
\put(1.9,4){UNSAT}

\put(7.2,2.8){RS sat/unsat w.~$\ham$}
\put(7.2,2.4){RS sat/unsat w.~$\ham'$}
\put(7.2,2.0){RS instability~$\ham$}

\end{picture}
\end{center}
\end{figure}

The line $\gamma^*_{\rm RS}(\epsilon)$, and even the ``triple point''
$\epsilon^*$ at which $\gamma^*_{\rm RS}(\epsilon)$ touches
$\gamma_{\rm uc}$, depend on the choice of Hamiltonian $\ham$
(\ref{eq.ham}) or $\ham'$ (\ref{eq.ham'}), and are plotted in
fig.~\ref{fig_RS}.  Since the ground-state energy 
$E(\beta \to \infty)$ (\ref{energy_entropy}) is zero if and only if
$\mathbf{q} = \mathbf{0}$ we conclude that the line 
$\gamma^*_{\rm RS}$ for $\epsilon < \epsilon^*$, and $\gamma_{\rm UC}$
for $\epsilon \ge \epsilon^*$ is the replica-symmetric prediction for
the SAT/UNSAT threshold.

There are striking hints towards the badness of the replica-symmetric solution.
For example, the
predicted critical value $\gamma^*_{\rm RS}(\epsilon=0)$ is different from the
numerical one, and even larger than the rigorous upper bound. But
there is also an inconsistency internal to the method. 
It is not possible that the satisfiability line in the phase diagram
depends on the finite-temperature Hamiltonian used in the cavity
equations, but the lines $\gamma^*_{\rm RS}$ coming from $\ham$ 
and $\ham'$ are different.
Another 
argument comes from the cavity prediction of the ground-state energy 
$E_{\textrm{min}}=E(\beta \to \infty)$ (\ref{energy_entropy})
which is zero if and only if $\mathbf{q} = \mathbf{0}$. The
\emph{discontinuous} appearance of a new fixed point leads to a
discontinuity in $\gamma$ of 
$ E_{\textrm{min}}(\gamma, \epsilon)$, but this is impossible,
as we have the Lipschitz condition
\be
\deenne{\gamma}{{}}
E_{\textrm{min}}(\gamma, \epsilon)
\in \left[ 0, \smfrac{1}{3} \right]
\ee
coming from the fact that adding randomly $M'$ clauses to an instance
whose minimum energy is $E_{\textrm{min}}$ can only give an instance
with minimum energy in the range 
$[E_{\textrm{min}}, E_{\textrm{min}} + M' ]$.
In section \ref{RS_stab} we will explain in detail why and where exactly 
the replica-symmetric solution breaks down.

\subsection{The soft-fields analysis}
\label{entropic_cav}

In the region of the phase diagram where the RS hard-fields analysis 
predicts zero ground-state energy (SAT region) all 
the hard fields and biases are zero ($g_+(0), g_-(0) > 0$ in the
language of equation (\ref{eq.lowTfields})). We denote with capital
letters $U_{a \to i}$ and $H_{i \to a}$ the soft fields, i.e.
\begin{align}
u_{a \to i} &= 0 + \frac{U_{a \to i}}{\beta}
\ef;
&
h_{i \to a} &= 0 + \frac{H_{i \to a}}{\beta}
\ef.
\end{align}
Their update is deduced analyzing
the general cavity equations (\ref{BP_T})
\begin{align}
H_{i \to a} &= \sum_{b \in \di \setminus a} U_{b \to i} 
\ef;
\\
J_{ai} U_{a \to i} &
= -\frac{1}{2}\ln \left( e^{2 J_{aj} H_{j \to a}} 
  + e^{2 J_{ak} H_{k \to a}} \right)
\ef;
\label{BP_ent}
\end{align}
Defining $q(U)$ and $p(H)$ as the probability distributions of $U$ and $H$ over
the graph, we have the self-consistent equations
\begin{align}
p(H) &= (1-\epsilon) \tilde{p}(H) + \epsilon \,\tilde{p}(-H)
\ef;
\\
\tilde{p}(H) 
&= \sum_{k=0}^\infty e^{-\gamma} \frac{\gamma^k}{k!} 
\int \prod_{i=1}^k \big[ \dx{U_i} \, q(U_i) \big]
\; \delta \Big( H-\sum_{i=1}^k U_i \Big)
\ef;
\end{align}
being the analogue of (\ref{eq.WP1b}), while for the
(\ref{eq.WP1a}) one has
\begin{align}
q(U) &= (1-\epsilon) \tilde{q}(U) + \epsilon \,\tilde{q}(-U)
\\
\tilde{q}(U) &= \int \dx{H_j} \dx{H_k}
\; p(H_j) \; p(H_k) \;
\delta \Big( U + \frac{1}{2} \ln \big( e^{2 H_j} + e^{2 H_k} \big)
\Big)
\end{align}
These equations are already beyond the possibilities of an analytical
treatment, and can be solved only by a population-dynamics
technique~\cite{MP99}. 

In this ``paramagnetic'' region the expression for the RS
ground-state entropy (logarithm of the number of SAT configurations) 
simplifies. Thus, knowing the distributions $q(w)$ and $p(g)$ we can compute 
the average entropy. In appendix \ref{app_1st} we compute the average 
number of solutions $\langle {\cal N} \rangle$, annealed entropy 
$\ln(\langle {\cal N} \rangle)$, whereas the RS computation leads 
to quenched average of the entropy $\langle \ln{\cal N} \rangle$. 
The same quantity may be computed for a given large sparse graph from 
the message passing procedure (\ref{BP_ent}). However, the result will 
be valid only in the region where the RS assumption is valid, see
section~\ref{RS_stab}.

\subsection{The RS stability}
\label{RS_stab}

The replica-symmetric solution will turn out to be incorrect if
the assumption of having a single pure phase is proven to fail.
As we know, a necessary condition for this is that fields incoming to a
given node are uncorrelated. This property can be tested on the
RS solution: if the spin-glass susceptibility
diverges, then all the fields (and in particular
the pairs of incoming ones) are strongly co-fluctuating, and the RS assumption
is inconsistent.
The (nonlinear) spin-glass susceptibility is defined as \cite{MR03, RBMM04} 
\begin{align}
\label{chi2}
\chi_{\textrm{SG}}(G)
&=
\frac{1}{N} \sum_{i, j \in V(G)}
\langle s_i s_j \rangle^2_c
\ef;
&
\chi_{\textrm{SG}}
&=
\sum_{d=0}^\infty (2 \gamma)^d \;
{\mathbb{E}}(\langle s_0 s_d \rangle^2_c) 
\ef.
\end{align}
On the left is the definition for a fixed graph $G$,
$\langle s_i s_j \rangle_c$ is the connected correlation
function between nodes $i$ and $j$. On the right we consider the
average over instances, in the thermodynamic limit, where sites $s_0$
and $s_d$ are at distance $d$. The factor $(2 \gamma)^d$ stands for
the average number of neighbours at distance $d$, when $d \ll \ln N$.
Assuming that the limit for large $d$ of the summands in (\ref{chi2})
exists (with the limit $N \to \infty$ performed first), we relate it
to the \emph{stability parameter}:
\begin{equation}
   \overline \lambda = \lim_{d\to \infty}
(2 \gamma) \Big(
   {\mathbb{E}}(\langle s_0 s_d \rangle^2_c)
   \Big)^{\frac{1}{d}}
\ef.
\label{stab}
\end{equation}
Then the series in (\ref{chi2}) is essentially geometric, and converges if
and only if $\overline{\lambda}<1$.
         
Using the fluctuation-dissipation theorem we relate the
correlation $\langle s_0 s_d \rangle_c$ to the variation of
magnetization in $s_0$, caused by an infinitesimal magnetic field in
$s_d$. Then, one relates this quantity to cavity fields, i.e., up
to a factor ${\cal C}$ independent from $d$,
\begin{equation}
{\mathbb{E}}(\langle s_0 s_d \rangle^2_c) 
=
{\cal C} \cdot
\sum_{\substack{
a \in \partial d \\
b \in \partial 0 }}
{\mathbb{E}}
\left[\left( \frac{\partial u_{a \to d} }
  { \partial u_{b \to 0} }\right)^2 \right]
\ef.
\label{fdt0}
\end{equation}
Finally, using the fact that we perform the large-$N$ limit first, 
the variation above is dominated 
by the direct influence through the length-$d$ path between the two nodes,
and this induces a ``chain'' relation: if the path involves the
clause and variable nodes
$
(a_{d}, d, a_{d-1}, d-1, \ldots, a_0, 0)
$
we have
\begin{equation}
{\mathbb{E}}(\langle s_0 s_d \rangle^2_c)
=
{\cal C} \cdot
{\mathbb{E}}
    \left[ \prod_{\ell=1}^d
\left( \frac{\partial u_{a_{\ell} \to \ell} } 
    {\partial u_{a_{\ell-1} \to \ell-1} } \right)^2 \right]
\ef.
\label{fdt}
\end{equation}

\[
\setlength{\unitlength}{50pt}
\begin{picture}(3.5,2.4)(-1.2,-0.6)
\put(-0.7,-0.6){\includegraphics[scale=1, bb=200 310 335 430, clip=true]
{fig_smallFG.eps}}
\put(2,0.45){$i$}
\put(1.15,0.35){$a$}
\put(0.075,0.6){$j$}
\put(0.05,0.25){$k$}
\put(0.5,0.){$h_{k\to a}$}
\put(0.5,0.95){$h_{j\to a}$}
\put(1.18,0.78){$u_{a\to i}$}
\put(0.45,1.25){$\{ u_{b\to j} \}$}
\put(0.25,-0.3){$\{ u_{c\to k} \}$}
\put(-1.55,-0.3){$\{ c \in \partial k \setminus a \}$}
\put(-1.5,1.35){$\{ b \in \partial j \setminus a \}$}
\end{picture}
\]
%

\noindent
Inside the paramagnetic phase and in the zero-temperature limit (but
keeping soft fields of section \ref{entropic_cav}),
from (\ref{BP_ent}), we have for a path like the upper one in the
above figure 
\begin{equation}
  J_{ai} \frac{\partial u_{a \to i} } {\partial u_{b \to j}}  = J_{ai}
  \frac{\partial U_{a \to i} } {\partial U_{b \to j}} =
   \frac{-J_{aj} e^{2 J_{aj} H_{j \to a}} }{ e^{2J_{aj} H_{j\to a}}+
  e^{2J_{ak} H_{k\to a}}}
\ef.
\end{equation}
Instead of directly computing the stability parameter $\overline{\lambda}$, 
it is equivalent, and numerically easier,
to associate every bias $u_{a \to i}$ in the population-dynamics algorithm with a positive number $v_{a \to i}$. We update
this number together with the fields according to 
\begin{equation}
v_{a\to i}
= \sum_{b\in \djj \setminus a} 
  \left( \frac{\partial u_{a\to i} }{\partial  u_{b\to j}} \right)^2
  v_{b\to j}
 + \sum_{c\in \dk \setminus a} 
  \left( \frac{\partial u_{a\to i} }{\partial  u_{c\to k}} \right)^2
  v_{c\to k}
\ef,
\label{RS_MT}
\end{equation}
(we recall that, when performing population-dynamics technique,
the labels ``$a{}\to{}i$'' do not have any spatial
meaning: the population is just a collection, and messages are
randomly combined at each step).
After equilibration, the numbers $v$ will change on average
geometrically, with a factor $\overline{\lambda}$.

Numerically, we see that for a given $\epsilon$ the stability
parameter grows with connectivity $\gamma$. On fig.~\ref{fig_RS}
we see the line above which the replica-symmetric solution is unstable 
(i.e.~$\overline{\lambda}>1$). This line coincides with
the unit-clause upper bound within the errors from about
$\epsilon > 0.33 \pm 0.02$. In particular all the region where the 
RS results are contradictory is unstable. It is furthermore
remarkable (and unexpected) that there exist a region in which
RS solution is not stable, and yet the short clause heuristics
a.s.~finds a solution in polynomial time.

\section{1RSB cavity approach and its implications for the phase diagram}
\label{1RSB_cav}

The understanding of the role of ergodicity in the validity of the
replica-symmetric cavity assumption allows us to recast the cavity
method as a more powerful tool, for the case in which there 
are (exponentially) many phases. The assumptions underlying this process
go under the jargon term ``1RSB'' type of symmetry breaking 
\cite{MP99, MP03}. In the 1RSB approach we assume that exponentially many pure 
thermodynamical states (phases) exists, and that the neighbors 
of a node in absence of this node are uncorrelated only within 
each of these states. This happens because the cluster property (the small
correlation of observables far from each other) holds only within 
a pure phase.
The name replica symmetry breaking is due to 
historical reasons, since the mechanism was first proposed by Parisi
\cite{Parisi80}, while using the ``replica trick'' in analysing
a spin-glass model.

The necessary but conceptually impossible handling of the multiplicity
of pure phases may be replaced by a ``survey'' over these phases.  The
only memory of the original structure is through the free energies
$F_{\alpha}$ of the various phases $\alpha$. As the phases have to be
weighted with a Boltzmann weight in $F_{\alpha}$, a 
\emph{reweighting term} has to be introduced in the ``survey''
equations, as we show in section \ref{G1RSBce}.
In the zero-temperature limit the analysis leads to what are now called 
\emph{survey propagation equations}, which are developed fully in
section~\ref{ytoinfty}.

The solution to these equations is described in section \ref{1RSBresP1},
in section \ref{1RSBresP2} are results of the stability analysis
of this solution.
Checking the validity of the 1RSB cavity assumption (1RSB stability
analysis) gives us a hint if this solution could be the final correct
one.  This has been done for the $K$-SAT \cite{MPR03,MMZ05} and
Coloring problems \cite{KPW04} on random graphs. The results in those
cases supported strongly the conclusion that the SAT/UNSAT thresholds
computed with the 1RSB cavity method were exact.
The 1RSB stability analysis is technically involved, we summarize 
the main steps in appendix~\ref{S1RSB}.

\subsection{General 1RSB cavity equations}
\label{G1RSBce}

We define a complexity function $\Sigma(F)$ as the logarithm of
number of states, and hence it is computable by Legendre transform  
\begin{align}
-\beta m \Phi(m,\beta) 
&=
- \beta m F(\beta) + \Sigma(F)
\ef;
&
\frac{\partial \Sigma(F)}{\partial F}
&=
\beta m
\ef;
\end{align}
where the parameter $m$ plays role of a second temperature, for free
energies of states instead of energies of configurations,
and is called 
the Parisi parameter of the replica symmetry breaking. 
The function $\Phi(m,\beta)$
is called the ``replicated free energy''. 

Instead of one field and one bias on every edge, now we need to keep one
field and one bias for every edge \emph{and} every state. Or
equivalently, as we assumed that there is a huge number of states,
a distribution of fields and biases on every edge. 
The self-consistent equation for this distribution is \cite{MP99, MP03} 
\be
\begin{split}
{\cal P}^{a \to i}(u_{a\to i}) 
&=
Z_{a\to i}^{-1}
\int \prod_{b\in \djj \setminus a} 
\dx{u_{b\to j}} {\cal P}^{b \to j}(u_{b\to j})
\int \prod_{b\in \dk \setminus a}
\dx{u_{b\to k}} {\cal P}^{b \to k}(u_{b\to k})   
\\
& \quad \cdot \ 
\delta \big( u_{a\to i} 
               - {\cal F}(\{ u_{b\to j} \},\{ u_{b\to j} \}) \big)
 \cdot \exp(-\beta m \Delta F^{a \to i})
\ef.
\label{1RSB}
\end{split}
\ee
The function ${\cal F}$ is the single-phase
update of biases given by equations (\ref{BP_T}), 
and the last term is the reweighting of states, where 
$\Delta F^{a \to i}$ is the free-energy shift after adding clause $a$
and all its
neighbors except $i$. Referring to our calculations in equation
(\ref{free_s1}),
this free-energy shift is
\be
\begin{split}
e^{-\beta \Delta F^{a\to i} } 
&= 
\frac{Y_{+1}^{a\to i} + Y_{-1}^{a\to i}}
     {\prod_{j\in \da \setminus i} \prod_{b\in \djj \setminus a}
            ( Y_{+1}^{b\to j} +  Y_{-1}^{b\to j} ) }
\\
&=
\frac{ \sum_{s_i,s_j,s_k}
 \exp \big[ \beta \big( h_{j\to a}s_j + h_{k\to a}s_k 
             - E_{\{J_a\}}(s_i,s_j,s_k) \big) \big] }
     { \prod_{j\in \da \setminus i}
       \prod_{b\in \djj \setminus a} 
         ( 2 \cosh{\beta u_{b\to j}} )}
\ef.
\label{reweigth} 
\end{split}
\ee
Then, in analogy with (\ref{free}), the replicated free energy $\Phi$
is calculated as 
\begin{equation}
    \Phi(m,\beta) = \sum_a \Delta \Phi^{a \cup \da} - \sum_i (d_i-1) \Delta
    \Phi^{i}\ ,  \label{1RSB_free}
\end{equation}
where $d_i$ is the degree of the node $i$.
The replicated free-energy shifts are
\begin{align}
e^{-\beta m \Delta \Phi^{a \cup \da} } 
&=
\int e^{-\beta m \Delta F^{a \cup \da} }
\ef;
&
e^{-\beta m \Delta \Phi^{i} }
&=
\int  e^{-\beta m \Delta F^{i} }
\ef.
\label{phi_s}
\end{align}
The integrals are making an average over the distributions of the
fields incoming to the cavity, similarly as in equation~(\ref{1RSB}),
i.e.~$\int f(u_1, \ldots, u_k) = \int \dx{u_1} {\cal P}^{(1)}(u_1)
\cdots \int \dx{u_k} {\cal P}^{(k)}(u_k) f(u_1, \ldots, u_k)$.

Since we are interested mainly in the ground-state properties of the 
$\epsilon$--1-in-3 SAT
problem, we need to take the zero-temperature limit. There are two
standard ways of doing this
\begin{description}
   \item{\bf The energetic $T\to 0$ limit \cite{MP03}:} We take the limit $\beta \to
       \infty$, $m\to 0$ with $y=\beta m$ fixed and finite. 
       Then
\begin{equation}
 -y\Phi(y)=-yE+\Sigma(E)
\ef.
\label{Legr} 
\end{equation}
Here we neglected the entropic
       contribution and we can obtain complexity as a function of energy. 
       This is the 1RSB analog of the RS analysis with hard fields
       (warnings).
   \item{\bf The entropic $T\to 0$ limit \cite{MPR05}:} We take the limit $\beta \to
       \infty$ at energy fixed to zero,  $E=0$. Then
\begin{equation}
m \tilde{\Phi}(m)= m S+\Sigma(S)
\ef;
\label{Legr_entropic}
\end{equation}
where $-\beta\Phi(\beta,m)\to \tilde{\Phi}(m)$. Here we are
       fixed to zero energy, but on the other hand we can compute complexity
       (number of states) as a function of the state internal entropy. 
       This is the 1RSB analog of the RS analysis with soft fields (beliefs).
\end{description}
In this paper we work out only the simpler energetic limit, the same as in 
\cite{MZ02} for $K$-SAT, or in \cite{MPWZ02} for Coloring. 
We will see how this analysis alone already gives us a large amount of
information about the phase diagram.

The reweighting (\ref{reweigth}) becomes in the zero-temperature energetic 
limit
\be
\Delta E^{a\to i}
=
- \max_{s_i,s_j,s_k}
 \big( h_{j\to a} s_j + h_{k\to a} s_k
       - E_{\{J_a\}}(s_i,s_j,s_k) \big)
+ \sum_{b\in \djj \setminus a}  |u_{b\to j}|
+ \sum_{b\in \dk \setminus a}   |u_{b\to k}|
\ef.
\label{rew_0} 
\ee
Since the fields $h$ and biases $u$ are integers, by relations 
(\ref{eqs.cavityZT}), $\Delta E^{a\to i}$ also takes 
nonnegative integer values. In fact it counts the number of contradictions
in one warning-propagation update. 

We want to determine whether a typical $\epsilon$--1-in-3 SAT 
instance has any satisfying configuration.
We do this in section \ref{ytoinfty}, by taking the $y\to \infty$
limit. The reweighting term
$\exp{(-y \Delta E^{a\to i})}$ then guarantees that we keep only the cases
without contradictions, $\Delta E^{a\to i}=0$. 
Conversely, in order to compute the ground-state energy in the 
UNSAT region, or the complexity at energies higher than zero, we need to 
keep $y$ finite. We undertake this in appendix \ref{yfinite}.

\subsection{The zero-energy case, survey propagation}
\label{ytoinfty}

In the limit $y \to \infty$ we fix the energy to be zero. 
The energy shift $\Delta E^{a\to i}$ is zero if and only if:  
\begin{itemize}
   \item The $\{ u_{b\to j} \}_{b \in \djj \setminus a}$ 
         are all nonnegative or all nonpositive,
   \item The $\{ u_{b\to k} \}_{b \in \dk \setminus a}$
         are all nonnegative or all nonpositive, 
   \item The terms $J_{aj} \sum_{b\in \djj \setminus a} u_{b\to j}$
         and $J_{ak} \sum_{b\in \dk \setminus a} u_{b\to k}$
         are not both positive.
\end{itemize}
We begin by simplifying the form of eq.~(\ref{1RSB}). In the 
zero-temperature energetic limit we can write the distribution of
fields and biases over states on every edge as 
\begin{align}
{\cal P}^{a \to i}(u_{a\to i})
&=
  q_-^{a \to i} \delta(u_{a\to i} + 1)
+ q_+^{a \to i} \delta(u_{a\to i} - 1)
+ q_0^{a \to i} \delta(u_{a\to i} )
\ef;
\\ 
{\cal P}^{i \to a}(h_{i\to a})
&=
  p_-^{i \to a} \mu_-(h_{i\to a})
+ p_+^{i \to a} \mu_+(h_{i\to a})
+ p_0^{i \to a} \delta(h_{i\to a})
\ef;
\end{align}
where
$q_-^{a \to i} + q_+^{a \to i} + q_0^{a \to i} 
= p_-^{i \to a}+p_+^{i \to a} +p_0^{i \to a}=1$, and
$\mu_{\pm}(h)$ are normalized measures with support over
$\mathbb{Z}^{\pm}$.

So, to every oriented edge we associate a triple of numbers 
$q=(q_-,q_0,q_+)$ or a triple $p=(p_-,p_0,p_+)$
(resp.~if oriented towards the variable or the clause).
In analogy with the self-consistent equations (\ref{BP_T}) for fields 
and biases we can write self-consistent equations for probabilities 
(surveys) $q$ and $p$.

\[
\setlength{\unitlength}{50pt}
\begin{picture}(3.5,2.4)(-1.2,-0.6)
\put(-0.7,-0.6){\includegraphics[scale=1, bb=200 310 335 430, clip=true]
{fig_smallFG.eps}}
\put(2,0.45){$i$}
\put(1.15,0.35){$a$}
\put(0.075,0.6){$j$}
\put(0.05,0.25){$k$}
\put(0.5,0.){$p^{k\to a}$}
\put(0.5,0.95){$p^{j\to a}$}
\put(1.18,0.78){$q^{a\to i}$}
\put(0.45,1.25){$ q^{b\to j} $}
\put(0.25,-0.3){$ q^{c\to k} $}
\put(-1.55,-0.3){$\{ c \in \partial k \setminus a \}$}
\put(-1.5,1.35){$\{ b \in \partial j \setminus a \}$}
\end{picture}
\]

\noindent 
Considering only the combinations with $\Delta E^{a\to i}=0$, the surveys 
of fields are given by incoming surveys of biases as
\begin{subequations}
\label{SP_1}
\begin{align}
p_+^{i \to a} + p_0^{i \to a}
  &=
  {\cal N}_{i \to a}^{-1}
  \prod_{b\in \di \setminus a} ( q_+^{a \to i}  + q_0^{a \to i} )
\ef;
\\
p_-^{i \to a} + p_0^{i \to a}
  &=
  {\cal N}_{i \to a}^{-1}
  \prod_{b\in \di \setminus a} ( q_-^{b \to i}  + q_0^{b \to i} )
\ef;
\\
p_0^{i \to a}
  &=
  {\cal N}_{i \to a}^{-1}
  \prod_{b\in \di \setminus  a} q_0^{b \to i}
\ef;
\end{align}
\end{subequations}
where ${\cal N}_{i \to a}$ is the normalization factor
(in the update, we have three equations for three independent unknown,
$p_{\pm}$ and $\mathcal{N}$).
And the surveys of biases are given by the incoming surveys 
of fields
\begin{subequations}
\label{SP_2}
\begin{align}
q_{J_{ai}}^{a \to i}
  &=
  {\cal N}_{a \to i}^{-1} \; 
  p^{j \to a}_{-J_{aj}} p^{k \to a}_{-J_{ak}}
\ef;
\\
q_{-J_{ai}}^{a \to i}
  &=
  {\cal N}_{a \to i}^{-1} \; 
  \big( p_{J_{aj}}^{j \to a} (1-p_{J_{ak}}^{k \to a} )
        + (1-p_{J_{aj}}^{j \to a} ) p_{J_{ak}}^{k \to a} \big)
\ef;
\\
q_{0}^{a \to i}
  &=
  {\cal N}_{a \to i}^{-1} \;
  \big( p_{-J_{aj}}^{j \to a} p_{0}^{k \to a} 
      + p_{0}^{j \to a} p_{-J_{ak}}^{k \to a}
      + p_{0}^{j \to a} p_{0}^{k \to a}  \big)
\ef;
\end{align}
\end{subequations}
where ${\cal N}_{a \to i}$ is the normalization factor, and the lower indexes
of $q$'s and $p$'s are multiplied by $-1$ when variable is negated in the 
clause.   

These equations describe survey propagation \cite{MZ02}, 
and can be used inside an algorithm to find a solution to a
typical instance of $\epsilon$--1-in-3 SAT, hopefully also in a region
of parameters $(\epsilon, \gamma)$ where short-clause--like heuristics
or belief-propagation methods fail.
They might also be used to compute quantities averaged over instances
in our random ensemble, particularly the complexity function which 
determines the SAT/UNSAT transition.

In the zero-temperature limit, the replicated free energy
(\ref{1RSB_free}) becomes
\begin{equation}
 -y \Phi(y)
 = \sum_a \ln \left( \int e^{-y \Delta E^{a \cup \da}} \right)
 - \sum_i (d_i-1) \ln \left( \int  e^{-y \Delta E^{i} } \right)
\ef,
\label{1RSB_free_0}
\end{equation}
where from (\ref{free_s}) and (\ref{phi_s}) we get the energy shifts 
\begin{align}
 \Delta E^{a \cup \da}
&=
 - \max_{s_i,s_j,s_k}
 \big( h_{i\to a} s_i + h_{j\to a} s_j + h_{k\to a} s_k 
      - E_{\{J_a\}}(s_i,s_j,s_k) \big)
 + \sum_{i\in \da} \sum_{b\in \di \setminus a} |u_{b\to i}|
\ef;
\\ 
 \Delta E^{i}
&=
 - \Big| \sum_{a\in \di} u_{a\to i} \Big|
 + \sum_{a\in \di} |u_{a\to i}|
\ef;
\label{en_s}
\end{align}
again both these energy shifts are non-negative integers.

Furthermore in the $y \to \infty$ limit, we distinguish only if $\Delta E=0$,
then $\exp(-y \Delta E)=1$, or if $\Delta E>0$, then $\exp(-y \Delta E)=0$. 
From eq.~(\ref{1RSB_free_0}) we get for the complexity at zero energy
\begin{equation}
   \Sigma(E=0)
 = \sum_a \ln \big( \textrm{prob}(\Delta E^{a \cup \da}=0) \big)
 - \sum_i (d_i-1) \ln \big( \textrm{prob}(\Delta E^{i}=0)  \big)
\ef,
\label{Sigma_0}
\end{equation}
where, calling
${\cal P}_0^i := \prod_{a\in \di} q_0^{a \to i}$ and
${\cal P}_{\pm}^i := \prod_{a\in \di} ( q_{\pm}^{a \to i}
   + q_0^{a \to i} )$,
\begin{subequations}
\label{eq.prob_12}
\begin{align}
\textrm{prob}(\Delta E^{i}=0)
&=
{\cal P}_+^i + {\cal P}_-^i - {\cal P}_0^i
\ef;
\label{prob_1}
\\
\begin{split}
\textrm{prob}(\Delta E^{a \cup \da}=0)
&=
\prod_{i\in \da} ({\cal P}_{J_{ai}}^{i\to a} 
   + {\cal P}_{-J_{ai}}^{i\to a} - {\cal P}_0^{i\to a} )
-\prod_{i\in \da} ({\cal P}_{-J_{ai}}^{i\to a} - {\cal P}_0^{i\to a} )
-\prod_{i\in \da} ({\cal P}_{ J_{ai}}^{i\to a} - {\cal P}_0^{i\to a} )
\\
& \quad
-\sum_{i \in \da}
{\cal P}_{-J_{ai}}^{i \to a}
\prod_{j \in \da \setminus i} 
      ({\cal P}_{J_{aj}}^{j \to a} - {\cal P}_0^{j \to a} )
\ef.
\end{split}
\label{prob_2}  
\end{align}
\end{subequations}
The second equation collects the contributions from
all combinations of arriving fields except the ``contradictory'' ones
$(+,+,+)$, $(-,-,-)$, $(+,+,0)$ and $(+,+,-)$ (plus permutations of
the latters). Note at this point that all these equations (\ref{SP_1})-(\ref{eq.prob_12}) correctly do not depend 
on the choice of Hamiltonian (\ref{eq.ham}) or (\ref{eq.ham'}).

\subsection{1RSB results for the phase diagram}
\label{1RSBres}

In this section we give results of 1RSB cavity analysis for the 
$\epsilon$-1-in-3 SAT problem. In the first two subsections
we concetrate on the positive 1-in-3 SAT ($\epsilon=0$).
In the third one \ref{1RSBreseps} we show results for general 
probability of negation.

\subsubsection{Complexity as a function of connectivity}
\label{1RSBresP1}

\begin{figure}[!ht]
\begin{center}
\setlength{\unitlength}{50pt}
\begin{picture}(6,3)
\put(0,0){\includegraphics[scale=0.65]{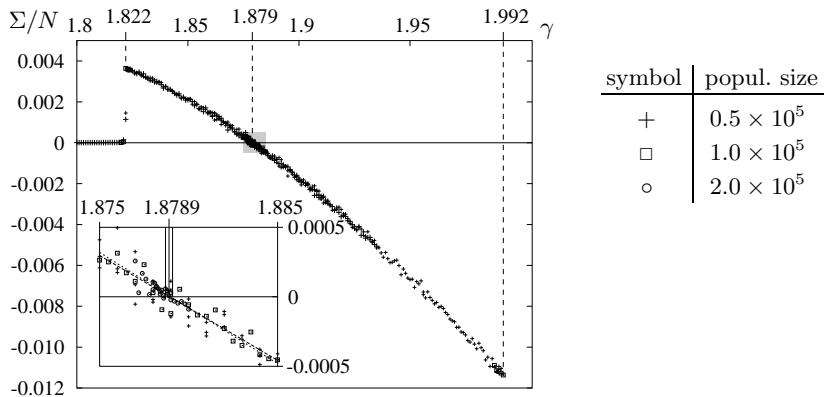}}
\put(0.08,2.85){$\Sigma/N$}
\put(4.1,2.8){$\gamma$}
\put(4.5,2){
\begin{tabular}{c|c}
symbol\rule{2pt}{0pt}& \;popul.~size \\
\hline
$+$                     & $0.5 \times 10^5$\rule{0pt}{12pt}
\\
$\scriptstyle{\square}$ & $1.0 \times 10^5$ \\
$\circ$                 & $2.0 \times 10^5$ \\
\end{tabular}
}
\end{picture}
\caption{\label{sigma}%
Average complexity density (logarithm of number of states divided by size 
of the graph) 
as a function of mean degree $\gamma$ for the positive 1-in-3 SAT problem.
At $\gamma_{\rm sp}=1.822$
a nontrivial solution of survey propagation equations appears,
with positive complexity.
At $\gamma=\gamma^*=1.8789\pm 0.0002$ the complexity becomes negative: 
this is the SAT/UNSAT transition. At $\gamma_p=1.992$ the solution 
at zero energy ceases to exist.
The inset magnifies the region where the
complexity crosses zero, together with the error bar for the SAT/UNSAT 
transition.
}
\end{center}
\end{figure}

To compute numerically the average value of complexity from
eq.~(\ref{Sigma_0}) we first need to find a fixed point of the survey
propagation equations (\ref{SP_1}) and (\ref{SP_2}). We do that using
the population-dynamics algorithm \cite{MP99}. The result is in
fig.~\ref{sigma}.

Below mean degree $\gamma_{\rm sp}=1.822\pm 0.001$ 
the survey propagation equations (\ref{SP_1}, \ref{SP_2}) have only the
trivial paramagnetic solution, with
$p_0=q_0=1$ and $q_{\pm}=p_{\pm}=0$ for all edges.
At $\gamma_{\rm sp}$ a solution of 
survey propagation equations with positive complexity appears
discontinuously.
The emergence of this transition far below the numerically known 
SAT/UNSAT threshold suggests that, in a whole interval of parameters
near to the threshold 
the phase space restricted to solutions is clustered into many pure
states, a {\it Hard-SAT phase} \cite{MZ02} exists. 
Furthermore, in that interval, there 
are also many metastable states, entropically relevant,
and local minima with positive energy cost
separated by macroscopic barriers.
This means that local algorithms, like 
decimation heuristics or variants of
annealed Monte Carlo, get trapped and are unable to find any ground
state in polynomial time.
Nonetheless, a decimation procedure based on the stationary
distribution of survey propagation equations is expected to work
beyond this threshold.

Note that $\gamma_{\rm sp}$ is referred to as a ``dynamical 
threshold'' in \cite{MZ02,MPWZ02}, we stress that it is not this 
point which is connected to a real dynamical transition \cite{us}. 
Neither is it the point where the local algorithm ceases to work in
polynomial time. We come back to comment about this point in the
discussion, section~\ref{discussion}.

At mean degree $\gamma^*=1.8789\pm 0.0002$ the complexity becomes
negative. Instead of having a.s.~in each instance an exponential
number of clusters which contain at least one solution, the fraction
of instances having any cluster which contains at least one solution
(i.e.~the fraction of SAT instances)
becomes exponentially small. So this point identifies the SAT/UNSAT
transition.

We are aware of two works where results from numerical
simulation for this SAT/UNSAT threshold are given. In \cite{MK05} 
they conclude the value of threshold is $\gamma^* = 1.86\pm 0.03$. 
In \cite{KSM04} they give $\gamma^* = 1.875\pm 0.015$. In fact, the 
latter do not give an error bar, so we guessed it from their fig.~4. 
Our result agrees with these estimations, and as it is based on 
an analytical method we reduce the error bar by one order of magnitude 
with a very small numerical effort.

At mean degree $\gamma_p=1.992\pm 0.001$ the solution at zero energy
ceases to exist.  In the $y \to \infty$ limit the population dynamics
converges to a solution which shows a finite fraction of surveys of
type $(p_{\pm},p_0,p_{\mp})=(0,0,1)$. Then, with finite probability we
would find two such surveys creating a contradiction, the
normalization in (\ref{SP_2}) then would be zero. We call this
situation a ``hard contradiction''.

Note that such a phenomenon does not occur in $K$-SAT or Coloring
problems. Cavity equations deal with the messages incoming to a clause
from all neighbours but one. In both $K$-SAT and Coloring (and in many
other problems, like NAE-$K$-SAT, Vertex Covering, and so on), there
is no way of making a clause unsatisfied if one of the neighbouring
variables is not restricted, and indeed a $\gamma_p$ threshold has
never appeared in the cavity analysis for these systems.

In order to obtain a non-singular solution above connectivity
$\gamma_p$, we need to work with the equations at finite $y$, which is
able to account for the positive energy contributions. The results
of the finite $y$ analysis are shown in appendix \ref{1RSBcomp}.

\subsubsection{The stability analysis} 
\label{1RSBresP2}

\begin{figure}[!ht]
\setlength{\unitlength}{50pt}
\begin{picture}(3.1, 2.85)(0.4,-0.2)
 \put(0,0){\includegraphics[scale=0.5]{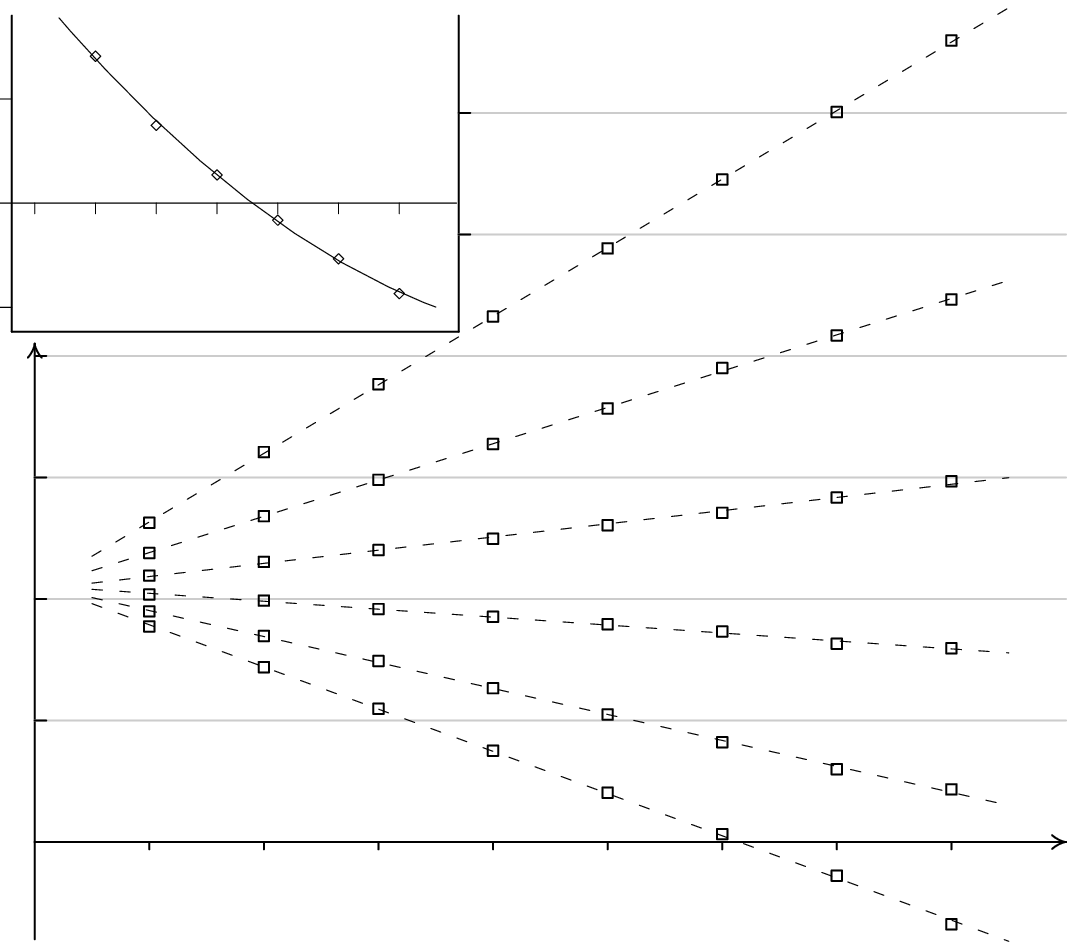}}
 \put(3.05,0.23){$d$}
 \put(-0.9,1.5){$d \ln \mu_{\rm II}(d)$}
 \put(-0.02,1){${\scriptstyle 0}$}
 \put(-0.14,1.35){${\scriptstyle 0.5}$}
 \put(-0.14,1.70){${\scriptstyle 1.0}$}
 \put(-0.25,0.65){${\scriptstyle -0.5}$}
 \put(-0.25,0.3){${\scriptstyle -1.0}$}
 \put(2.9,2.7){${\scriptstyle \gamma = 1.825}$}
 \put(2.9,1.9){${\scriptstyle \gamma = 1.83}$}
 \put(2.9,1.35){${\scriptstyle \gamma = 1.835}$}
 \put(2.9,0.8){${\scriptstyle \gamma = 1.84}$}
 \put(2.9,0.46){${\scriptstyle \gamma = 1.845}$}
 \put(2.9,0.){${\scriptstyle \gamma = 1.85}$}

 \put(-0.57,2.72){$\ln \mu_{\rm II}$}
 \put(1.,2.1){${\scriptstyle 1.85}$}
 \put(0.3,2.1){${\scriptstyle 1.83}$}
 \put(-0.05,2.5){\makebox[0pt][r]{${\scriptstyle 0.1}$}}
 \put(-0.05,2.2){\makebox[0pt][r]{${\scriptstyle 0}$}}
 \put(-0.05,1.9){\makebox[0pt][r]{${\scriptstyle -0.1}$}}

 \put(0.07,2.075){$\gamma$}


 \put(0.70,0.25){${\scriptstyle 4}$}
 \put(1.36,0.25){${\scriptstyle 8}$}
 \put(2.01,0.25){${\scriptstyle 12}$}
 \put(2.67,0.25){${\scriptstyle 16}$}

\end{picture}
\quad
\setlength{\unitlength}{40pt}
\begin{picture}(3.8,3.8)
\put(0,0){\includegraphics[scale=0.6]{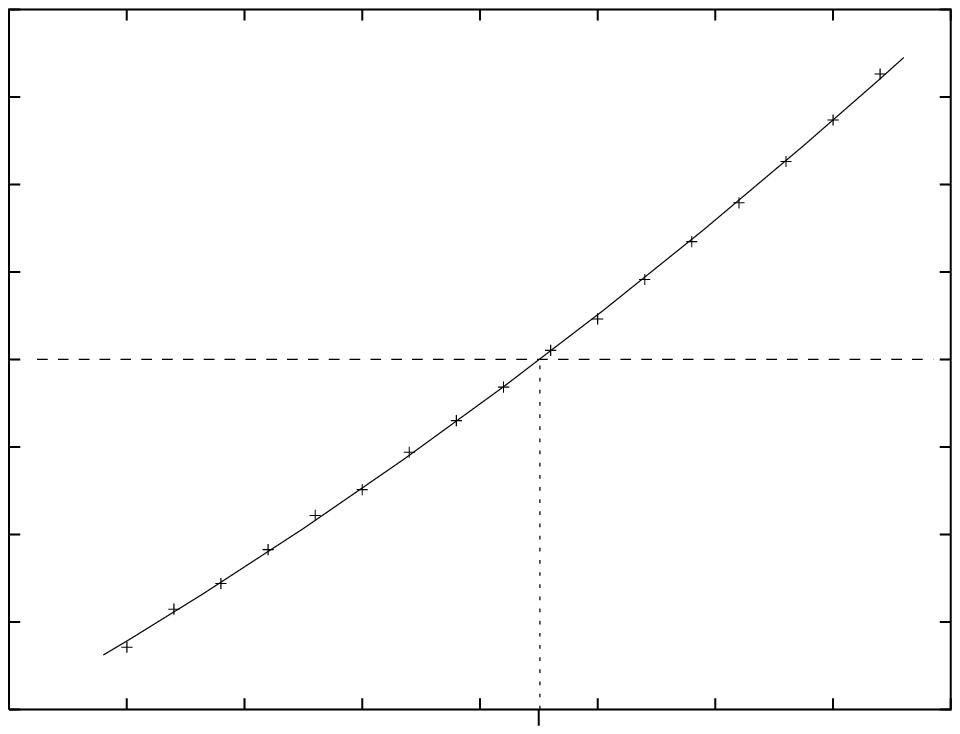}}
\put(4.82,0.45){$\gamma$}
\put(0.2,3.7){$\mu_{\rm I}$}

\put(4.10,0.3){${\scriptstyle 1.96}$}
\put(3.13,0.3){${\scriptstyle 1.95}$}
\put(2.03,0.3){${\scriptstyle 1.94}$}
\put(1.01,0.3){${\scriptstyle 1.93}$}
\put(2.58,0.22){${\scriptstyle 1.9475}$}

\put(0.255,0.42){${\scriptstyle 0.96}$}
\put(0.255,0.80){${\scriptstyle 0.97}$}
\put(0.255,1.18){${\scriptstyle 0.98}$}
\put(0.255,1.56){${\scriptstyle 0.99}$}
\put(0.255,1.94){${\scriptstyle 1   }$}
\put(0.255,2.31){${\scriptstyle 1.01}$}
\put(0.255,2.69){${\scriptstyle 1.02}$}
\put(0.255,3.07){${\scriptstyle 1.03}$}
\put(0.255,3.45){${\scriptstyle 1.04}$}
\end{picture}
\caption{\label{stabII} The stability parameter of the second kind 
$\ln(\mu_{\rm II}(d))$ (\ref{mu_d}) for positive 1-in-3 SAT for different 
connectivities as a function of length of the chain $d$. When the slope is 
negative, the 1RSB at this connectivity is stable against bug proliferation, 
and vice versa. This happens for $\gamma>\gamma_{\rm II}=1.838\pm0.002$.
Right: The stability parameter of the first kind $\mu_{\rm I}$ (\ref{lambda_I}) 
as a function of connectivity for positive 1-in-3 SAT. 
The stability parameter is smaller than 1 for 
$\gamma<\gamma_{\rm I}=1.948 \pm 0.002$,
for these connectivities the 1RSB equations are stable against noise
propagation.}
\end{figure}

In appendix \ref{S1RSB} we introduce two stability parameters $\mu_{\rm I}$
(\ref{lambda_I}) and  $\mu_{\rm II}$ (\ref{mu_d}). Their meaning is
analogous to that of the replica-symmetric stability parameter
$\overline \lambda$ 
(\ref{stab}). The 1RSB solution is stable if and only if both
$\mu_{\rm I}<1$ and $\mu_{\rm II}<1$. 

The results for stability parameter of the second kind $\mu_{\rm II}$
(\ref{mu_d}) for positive 1-in-3 SAT are shown in fig.~\ref{stabII}
(left), for finite $d$. Extrapolation to $d \to \infty$ is done by
linear fit, which looks reasonable from the data points. So our
criterion is that, if the slope in the logarithmic plot is positive, the
limit value $\mu_{\rm II}$ is larger than 1, and vice versa. We
estimate that 1RSB is ``type-II'' stable for 
$\gamma > \gamma_{\rm II} = 1.838 \pm 0.002$.

More directly, for the stability parameter of the first kind
$\mu_{\rm I}$ (\ref{lambda_I}), the results are shown in
fig.~\ref{stabII}, right. So we get that 1RSB is ``type-I'' stable for 
$\gamma < \gamma_{\rm I} = 1.948 \pm 0.002$ (the generous error
estimate is due to potential biases caused by finite population sizes).

The 1RSB solution may be correct only if both the stability parameters 
are smaller than one.
For positive 1-in-3 SAT this happens for connectivities in the range 
$1.838 \simeq \gamma_{\rm II} < \gamma < \gamma_{\rm I} \simeq 1.948$. 
So, in particular, the SAT/UNSAT threshold $\gamma^*$ is in the range
of stability, and
its value is to be considered exact.
Note that such a situation, in which the SAT/UNSAT
threshold falls into a narrow stability region, is quite common, and
it has been seen also in $K$-SAT \cite{MPR03} and Coloring~\cite{KPW04}.

\subsubsection{1RSB results for general probability of negation}
\label{1RSBreseps}

\begin{figure}[!ht]
\begin{center}
\setlength{\unitlength}{30pt}
\begin{picture}(12.2,6.4)(0,-1.)
\put(-0.12,-1.3){\includegraphics[scale=1., bb=0 0 270 250]
   {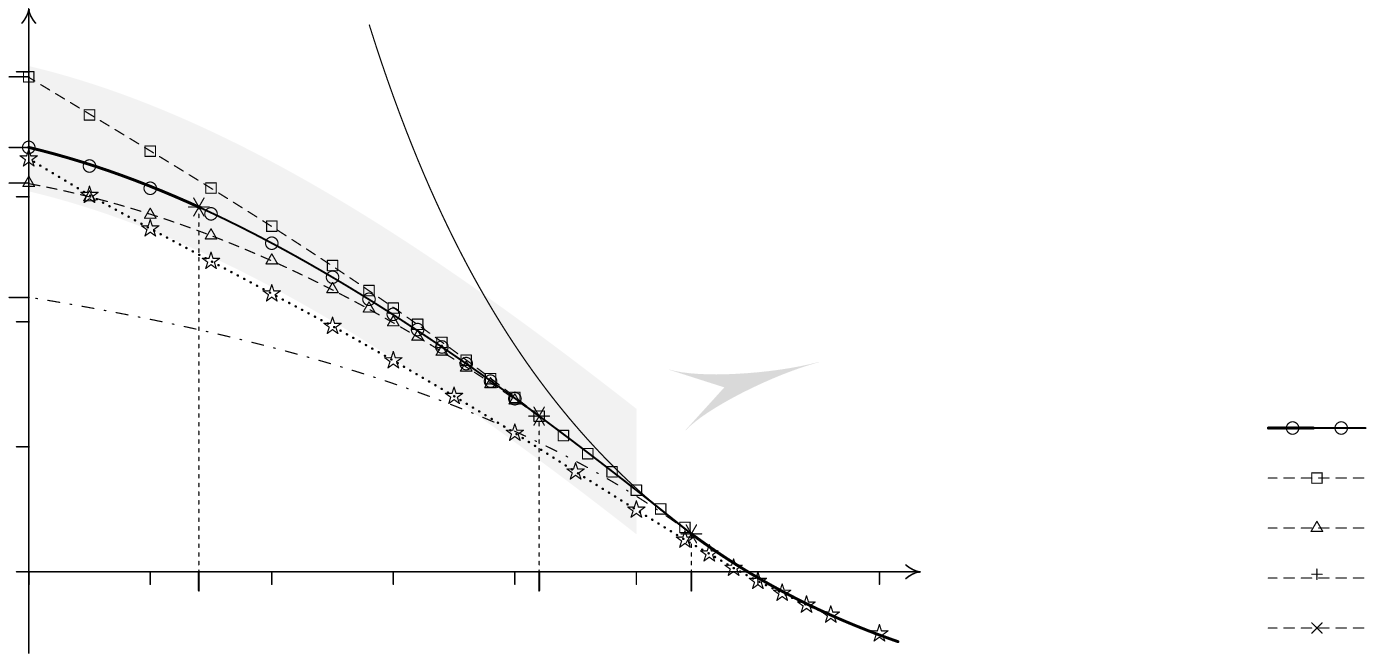}}
\put(8,2.5){\includegraphics[scale=1., bb=0 0 270 250]
   {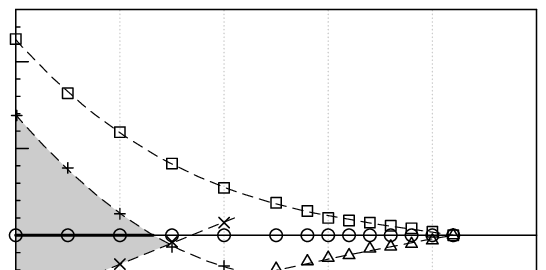}}
\put(1.166,-0.65){${\scriptstyle 0.05}$}
\put(2.393,-0.65){${\scriptstyle 0.1}$}
\put(3.500,-0.65){${\scriptstyle 0.15}$}
\put(4.676,-0.65){${\scriptstyle 0.2}$}
\put(5.833,-0.65){${\scriptstyle 0.25}$}
\put(7.060,-0.65){${\scriptstyle 0.3}$}
\put(8.166,-0.65){${\scriptstyle 0.35}$}
\put(1.65 ,-0.75){${\scriptstyle 0.07}$}
\put(5.05 ,-0.75){${\scriptstyle 0.21}$}
\put(6.32 ,-0.75){${\scriptstyle 0.2726}$}
\put(8.9  ,-0.6 ){$\epsilon$}
\put(-0.,4.85) {\makebox[0pt][r]{$\gamma$}}
\put(-0.1,-0.4){\makebox[0pt][r]{${\scriptstyle 1.2}$}}
\put(-0.1,0.8) {\makebox[0pt][r]{${\scriptstyle 1.4}$}}
\put(-0.1,2.0) {\makebox[0pt][r]{${\scriptstyle 1.6}$}}
\put(-0.1,3.17){\makebox[0pt][r]{${\scriptstyle 1.8}$}}
\put(-0.1,4.46){\makebox[0pt][r]{${\scriptstyle 2.0}$}}

\put(-0.1,2.246){\makebox[0pt][r]{${\scriptstyle \gamma_{\rm sch} = 1.639}$}}
\put(-0.1,3.69) {\makebox[0pt][r]{${\scriptstyle \gamma^* = 1.8789}$}}
\put(-0.1,3.39) {\makebox[0pt][r]{${\scriptstyle \gamma_{\rm sp} = 1.822}$}}
\put(-0.1,4.24) {\makebox[0pt][r]{${\scriptstyle \gamma_p = 1.992}$}}

\put(4.5,4){Easy UNSAT}
\put(2.2,0.6){Easy SAT}
\put(1.85,4.15){UNSAT}
\put(8.15,2.8){\makebox[0pt][r]{${\scriptstyle 0}$}}
\put(8.15,3.633){\makebox[0pt][r]{${\scriptstyle 0.05}$}}
\put(8.15,4.466){\makebox[0pt][r]{${\scriptstyle 0.10}$}}
\put(8.15,1.966){\makebox[0pt][r]{${\scriptstyle -0.05}$}}
\put( 9.1,1.8){${\scriptstyle 0.05}$}
\put(10.1,1.8){${\scriptstyle 0.10}$}
\put(11.1,1.8){${\scriptstyle 0.15}$}
\put(12.1,1.8){${\scriptstyle 0.20}$}
\put(13.1,2.55){$\epsilon$}
\put(8.15,5){\makebox[0pt][r]{$\gamma - \gamma^*(\epsilon)$}}

\put(12.7,1.3){\makebox[0pt][r]{\scriptsize{stable}
    \raisebox{-7.8pt}{\rule{0.25pt}{13pt}} }}
\put(11.9,0.95){\makebox[0pt][r]{$\gamma^*(\epsilon)$}}
\put(11.9,0.5){\makebox[0pt][r]{$\gamma_p(\epsilon)$}}
\put(11.9,0.05){\makebox[0pt][r]{$\gamma_{\rm sp}(\epsilon)$}}
\put(11.9,-0.45){\makebox[0pt][r]{$\gamma_{\rm I}(\epsilon)$}}
\put(11.9,-0.9){\makebox[0pt][r]{$\gamma_{\rm II}(\epsilon)$}}


\end{picture}
\end{center}
\caption{\label{fig_1RSB} Left: 
plot of the three curves 
$\gamma^*(\epsilon)$,
$\gamma_{\rm sp}(\epsilon)$ and
$\gamma_p(\epsilon)$ described in the text.
We left for comparison the SCH lower bound
(dot-dashed curve) and the RS instability line
(dotted curve with stars data points). 
Right: The same data, with connectivity plotted with 
   respect to the SAT/UNSAT threshold prediction
   $\gamma^*(\epsilon)$. Also the stability lines 
$\gamma_{\rm I}(\epsilon)$ and 
$\gamma_{\rm II}(\epsilon)$ are shown, and the interval of stability
   for the SAT/UNSAT curve is $\epsilon \in [0, 0.07 \pm 0.01]$.
In the inset, all the error bars are approximately as big as the point size.}
\end{figure}

We applied the techniques of section \ref{1RSBres} to the problem at
finite $\epsilon$. 
As expected, all the critical connectivities describe
curves which are continuous at $\epsilon=0$. We thus show, in
figure \ref{fig_1RSB} (left), the curves $\gamma^*(\epsilon)$,
$\gamma_{\rm sp}(\epsilon)$ and
$\gamma_p(\epsilon)$, and, in the magnification on the right
of fig. \ref{fig_1RSB}, also
$\gamma_{\rm I}(\epsilon)$ and 
$\gamma_{\rm II}(\epsilon)$.

As $\epsilon$ approaches about $0.20$, the interesting 
interval $\gamma_{\rm sp} < \gamma < \gamma_p$ becomes very narrow
(fig.~\ref{fig_1RSB}, right), 
and the complexity value very small, $\Sigma \approx 10^{-5}$,
three orders of magnitude smaller than the analogous values for $\epsilon=0$.
Above $\epsilon = 0.20$ we do not have sufficient numerical resolution 
to examine this region at all. 

In fig.~\ref{fig_1RSB} (right) 
we plot the four curves 
$\gamma_{\rm sp}(\epsilon)$, $\gamma_p(\epsilon)$,
$\gamma_{\rm I}(\epsilon)$ and 
$\gamma_{\rm II}(\epsilon)$, shifted by the curve
$\gamma^*(\epsilon)$, which is used as a reference.
This allows us to appreciate that 
the differences
$\big( \gamma^*-\gamma_{\rm sp} \big)(\epsilon)$ and 
$\big( \gamma_p-\gamma^* \big)(\epsilon)$ seem to vanish linearly at
about $\epsilon_p = 0.21 \pm 0.01$, these two linear fits
extrapolating to the same value of $\epsilon$ with reasonable
confidence.

Above $\epsilon_p$, as soon as a nontrivial solution of (energetic finite-$y$)
1RSB cavity equations exists, it has immediately a complexity
$\Sigma(E)$ of the qualitative shape for the connectivities above $\gamma_p$, 
(for example see fig.~\ref{sigm_ener} in appendix~\ref{yfinite}).
We are thus led to conclude that in this interval the line
$\gamma_p(\epsilon)$ should be taken as the 1RSB prediction for the
SAT/UNSAT line $\gamma^*(\epsilon)$.
At about $\epsilon=0.26 \pm 0.01$ the curve $\gamma_p(\epsilon)$
joints the unit-clause upper bound.

We should add that, above $\epsilon \simeq 0.07$, both stability
criteria fail along the curve $\gamma^*(\epsilon)$ (and then,
along $\gamma_p(\epsilon)$, for $\epsilon > \epsilon_p$), so that the 1RSB
prediction for the SAT/UNSAT transition is not expected to be exact,
but only an upper bound, in the range 
$0.07 < \epsilon < 0.2726$~\cite{FL03, FLT03}.

\section{Discussion and conclusions}
\label{discussion}

We studied the average-case behaviour of 1-in-3 SAT in the random
$\epsilon$--1-in-3 SAT ensemble, where $\epsilon \in [0, 1/2]$ is the
probability of negation. This generalizes the random (symmetric)
1-in-3 SAT problem ($\epsilon=1/2$) and random positive 1-in-3 SAT
problem ($\epsilon=0$), which is a special ensemble of Exact Cover.

Our main result is the phase diagram in fig.~\ref{fig_first}, and,
magnified, in fig.~\ref{fig_1RSB} above.
It fills the conceptual gap between the symmetric problem, of polynomial
average-case complexity both in the SAT and UNSAT regions,
and the positive problem, which shows a Hard-complexity phase around the
SAT/UNSAT threshold.

Concerning the SAT/UNSAT transition curve, $\gamma^*(\epsilon)$, we
computed upper bounds coming from unit-clause technique (UC), and from
first moment method with restriction to the 2-core (1MM), and the
lower bound coming from short clause heuristics (SCH).
The UC and SCH bounds have been proven to coincide on the interval 
$\epsilon \in [0.2726, 1/2]$, and thus determine the corresponding
portion of the SAT/UNSAT line rigorously.

All the other results are obtained with the non-rigorous cavity method. 
The results of the replica-symmetric calculations do not give us a 
better result for the SAT/UNSAT threshold than the UC and SCH bounds,
since its prediction have to be rejected above the RS stability line 
$\gamma_{\rm RS}$, fig.~\ref{fig_RS}.

It is remarkable that  
a region of the phase diagram exists, where the replica symmetry is broken, 
while the short clause heuristics is proven to succeed a.s.~in polynomial time.
For what we know, such a feature has not been proven in any of the 
previously studied models, while it has been often observed empirically
that some local algorithms, e.g. the Walk-SAT \cite{WS1,WS2,WS3,WS4}, 
works in linear time inside a phase with replica-symmetry breaking.
We are tempted to say that this result actually proves that the onset 
of a non-trivial replica symmetry broken solution does not imply 
to the onset of computational hardness (unfortunately the cavity results 
are not rigorous and the term ``computational hardness'' would have to be 
defined properly to be allowed to speak about a proof).
However, we hope that this could be used to study in a new way the 
nature of the replica-symmetry-broken phase.
On the other hand, and as claimed before, we are persuaded that a stable 
1RSB solution suggest an existence of a Hard-SAT phase near to the SAT/UNSAT
transition (nearer than the stability threshold). For quantitative study 
of this point for the coloring problem see  \cite{LenkaFlo}. The analysis
of \cite{LenkaFlo} should be repeated for 1-in-$K$ SAT problem in future works.

Our main insights for the region with $\epsilon<0.2726$ comes from the
one-step--Replica-Symmetry-Breaking (1RSB) calculations by analsis
of the energetic zero-temperature limit of the 1RSB cavity equations
(\ref{1RSB}), in which we keep only the weights of the hard fields instead
of whole probability distribution.

For $\epsilon<0.21$ we can locate, on the curves for the zero-energy
complexity function $\Sigma(\gamma)$ fig.~\ref{sigma}, the connectivities
$\gamma^*(\epsilon)$ at which $\Sigma$ vanishes, this corresponding to
the SAT/UNSAT transition.  The same computation shows the existence of
a nontrivial solution above 
$\gamma_{\rm sp}(\epsilon) < \gamma^*(\epsilon)$, thus predicting a whole
interval of Hard-SAT phase with many pure states. However for exact
location of the ``dynamical transition'' we would need to keep the information
about the soft fields and compute when a nontrivial solution of
eq.~(\ref{1RSB}) appears for the entropically dominating clusters, 
see~\cite{us}. 
Note here also that the inequality $\gamma_{\rm sp}<\gamma_{\rm RS}$ 
for small $\epsilon$ is due to the discontinuity of the transition 
towards nontrivial 1RSB phase, that can not be seen by the local 
RS stability analysis. Analysis of \cite{us} would also show that 
a nontrivial 1RSB solution exists everywhere in the region 
of $\gamma> \gamma_{\rm RS}$. This analysis for 1-in-$K$ SAT might be 
a direction for future work.

Above the line $\gamma_p(\epsilon)$, fig. \ref{fig_1RSB}, the system 
shows a transition to a
phase where the 1RSB solution at zero energy ceases to exists, while
it still exists above some value $E_{\rm min}(\gamma, \epsilon)$.
This is due to the presence of 
hard contradictions, a phenomenon specific of
strongly constrained problems, like 1-in-$K$ SAT, and to our knowledge
it is a newly observed fact.
Interpretation of this transition may be that the SAT formulas
start to be subexponentially rare at the connectivity
$\gamma_p$. 

For $\epsilon>0.21$ the nontrivial 1RSB solution at zero energy never
exists, and we can trace only the curve $\gamma_p$.  This would be the
1RSB prediction for the SAT/UNSAT transition.  This result suggests
that for $\epsilon>0.21$ a Hard-SAT region might actually be absent. 
Specifying what sort of replica-symmetry-broken solution 
is connected with the breakdown of local algorithms, like 
decimation heuristics or variants of annealed Monte Carlo,
is an important direction of future research. 

We have checked the local stability of 1RSB solution towards 2RSB. 
The result is that 1RSB is stable only in a small region between
the lines $\gamma_{\rm I}$ and $\gamma_{\rm II}$ in
fig.~\ref{fig_1RSB}. This means, among other things, that the 1RSB location 
of the SAT/UNSAT transition for $\epsilon<0.07$ is likely to be 
exact. In particular, this is true for the positive 1-in-3 SAT threshold,
$\gamma^* = 1.8789$. For $0.07 < \epsilon < 0.2726$ the 1RSB result 
is unstable so the exact location of the SAT/UNSAT transition in that region
remains an open question. 
We can only conjecture that our 1RSB result is an upper bound, in
analogy with proofs for other models in~\cite{FL03, FLT03}.

Furthermore, it would be interesting to compare our results with the
behaviour of the structurally affine $(2+p)$-SAT problem~\cite{MZ98,MZKST99} 
for which the 1RSB analysis is still missing.

Finally, as we mentioned in several places above, the 1RSB cavity approach 
allows for algorithmic implementations. We started studying this
aspect also together with Elitza Maneva and Talya Meltzer, and the
results will be published elsewhere.

\acknowledgments
We thank the organizers of the ``Complex Systems''
July~'06 Les Houches Summer School (Session 85), where this work has
been started. This work has been supported in part by the EU through
the network 
MTR 2002-00319 ``STIPCO'' and the FP6 IST consortium ``EVERGROW''.
A.S.~also thanks the LPTMS of Orsay -- Paris Sud for
support and kind hospitality in large stages of the work.

We thank also Elitza Maneva and Talya Meltzer for 
collaboration in some still unpublished algorithmic parts of this work.
We appreciated a lot discussions with Florent Krz\c{a}ka\l a,
R\'emi Monasson 
and last but not least we thank Marc M\'ezard for his precious hints.

\appendix

\section{Upper and Lower bounds from Unit-Clause Propagation
 and decimation heuristics}
\label{app_UC}

Consider an instance drawn from an appropriate ensemble, and subject to a decimation algorithm
wherby in each time set a variable is set to $\pm 1$.
Call $X$ the discrete decimation time (number of variables set, among
the $N$ total) and $C_i(X)$
the number of clauses remaining of length $i\geq 2$. Thus the initial conditions for
the instance are $X=0$ and $C_i(X)=N\frac{\gamma}{3}\delta_{i,3}$. Assume for
now these quantities are sufficient to describe the instance in the
absence of clauses of ``length 1'' (\emph{unit clauses}). If we assume a
variable is fixed (decimated) from such an instance, the remaining
instance involves a smaller number of literals, so it is simplified in
some respect. More importantly, some of the clauses are shortened, and
may even be reduced to unit clauses. The unit
clauses, being only 1 literal, allows no ambiguity in the values their
variables must take in order for the instance to be SAT. The initial
fixing of one variable by this process, forces the value of some other
variables, which may again propagate, i.e. a branching process.  So, a
single binary choice could decrease the number of variables by a
considerable amount, as a result of a cascade of these unit-clause
implications.

The justification in considering the instance at all times described
by $\{C_i(X)\}$ and $X$ is the following. For sufficiently simple
decimation rules, the distribution of remaining variables within
clauses will be uniform and random at all $X$, and 
if $N$ is sufficiently large the values of
$C_i(X)$ are self-averaging.  Furthermore, at fixed clause length, the
fraction of clauses with a given number of negations is the one
expected from an independent Bernoulli process: among clauses of
length $i$, at all times there is a fraction $\binom{i}{h} \epsilon^h
(1-\epsilon)^{i-h}$ of clauses with $h$ negations. All these elements
are necessary to allow a sufficiently concise dynamical description to
make the progress in the following sections. 
Among the various possible heuristics -- which determine the 
values set in the absence of unit clauses, one typically is interested in
the (suboptimal) subset of heuristics which preserve these decorrelation
properties of the Poissonian ensemble, so that a statistical analysis
is achievable.

It is useful to consider the algorithm as partitioned into rounds,
which consist of a single application of the heuristic rule (free
step), followed by the cascade of unit-clause propagations (forced
steps).
In expectation, the number of variables fixed throughout a  round of
unit-clause propagation
is described by a transition matrix depending
on the clause distribution $\{C_i(X)\}$, which are constant to leading
order during any ``subcritical'' round (defined below). The unit clauses generated in the
first free step go on to generate other unit clauses and so forth,
this can be described by a geometric series in the transition matrix
${\cal M}(X)$. Calling 
$\boldsymbol{p} = (p_T, p_F)$ and
$\boldsymbol{m} = (m_T, m_F)$ respectively the expectations for the
numbers of variables fixed to $({\rm True}, {\rm False})$ at the first
level of the cascade ($p$), and on the whole cascade ($m$), we have
\be
  \boldsymbol{m} = 
  \boldsymbol{p} + {\cal M}(x)\boldsymbol{p} 
  + {\cal M}^2(x)\boldsymbol{p} + \cdots 
  =
  (I-{\cal M}(x))^{-1} \boldsymbol{p}
\ef.
\label{geometric_series}
\ee
The matrix inverse above is justified from the fact that, for
consistency of the approximations, we require the round to be
subcritical, thus we must consider the range of parameters where the
modulus of the largest eigenvalue of ${\cal M}$ is smaller than
$1$. This is also responsible for the approximation that ${\cal M}(x)$
remains invariant (up to order $\frac{1}{N}$) during the cascade.

The transition matrix has two components. A first one comes from
$K$-clauses which are ``broken'' into a bunch of $K-1$ unit-clauses,
because the fixed variable already satisfies the original one, and all
the other $K-1$ variables thus should take the value not satisfying
the clause. A second one comes from 2-clauses (if any) which were just
``shortened'' because although the fixed variables are not satisfying them, 
the left over variables must, by definition creating unit clauses.  Since
there are $C_i(x)$ clauses of length $i$ and $N-X$ variables left in
the instance, these two terms are combined in the expression
\be
\label{eq.calM}
{\cal M}(x) = \frac{2}{N-X}
\left[
(C_2(X) + 3 C_3(X))
\begin{pmatrix}
\epsilon (1- \epsilon) & \epsilon^2 \\
(1-\epsilon)^2         & \epsilon (1- \epsilon)
\end{pmatrix}
+
C_2(X)
\begin{pmatrix}
\epsilon (1- \epsilon) & (1-\epsilon)^2 \\
\epsilon^2             & \epsilon (1- \epsilon)
\end{pmatrix}
\right]
\ef.
\ee
Here we identify that the unit-clause cascades on a large graph are a
simple uncorrelated process, governed by the spectrum of a certain
finite-size ``transition matrix'' (in our case, $2 \times 2$). If all
the eigenvalues have $|\lambda_i| < 1$, the process is
\emph{subcritical}: the typical size of the cascades is 
$\sim 1/\min_i (1-|\lambda_i|)$, and their average size concentrates.
Conversely, when the gap $1-|\lambda_i|$ vanishes, a single cascade
could visit a finite fraction of a graph even in the large $N$ limit,
and thus could lead to a contradiction.

Also note that both the upper- and lower-bound arguments later derived are
complemented by an analysis of the concentration properties of the
process~\cite{Deroulers:CUU}, and by a
non-rigorous argument on the approximate decorrelation of distinct
random restarts on a fixed instance. The vulgate version reads: ``If a
random algorithm succeeds with finite probability $p$ on its first
run, after $\sim n$ independent runs the probability of
success will be $\sim 1-\exp (-n \ln (1-p) + \cdots)$'', where the dots
stand for some function of the correlations, small in $n/N$, caused by
working with a fixed finite instance.  We do not discuss here these complex
technical points.

\subsection{Upper bound}
\label{ssec.UB}

If, for a variable $i$ selected randomly from the initial instance, 
both the cascades initiated by $s(i)=+1$ and $s(i)=-1$
percolate, there is a finite probability that they result in a
certificate of contradiction.

Thus the upper bound for the SAT/UNSAT transition 
comes from the requirement that the cascades are on
the edge of criticality already at time $X=0$. At this point we have
$C_2=0$ and $C_3= N \gamma/3$, so that we get
\be
{\cal M}(0) = 2 \gamma
\begin{pmatrix}
\epsilon (1- \epsilon) & \epsilon^2 \\
(1-\epsilon)^2         & \epsilon (1- \epsilon)
\end{pmatrix}
\qquad \longrightarrow \qquad
\big( \lambda_1, \lambda_2 \big) = 
\big( 4 \gamma\, \epsilon (1- \epsilon), 0 \big)
\ef.
\ee
From this we see that a random instance is a.s.~(randomized
linear time) provable to be unsatisfiable for $\gamma$ larger than the
percolation threshold
\be
\gamma_{\rm uc}(\epsilon) =
\frac{1}{4 \epsilon (1-\epsilon)} \label{A4}
\ef.
\ee

\subsection{Lower bound}
\label{ssec.LB}

The differential equations studied here are a generalization of those
found by Kalapala and Moore~\cite{MK05} for the case of
positive 1-in-$K$-SAT.

The heuristic determines the nature of the free step in our rounds.
The two rules examined here in are random heuristic (RH$[p]$) and
short clause heuristic (SCH). In RH$[p]$ a variable is chosen at
random, with uniform probability, from the remaining unassigned
variables and set to true with probability $p$. In SCH, a random
2-clause is selected (if any exists) and a random literal set to
satisfy the clause (hence the other literal is set to not satisfy it).
If at some time no 2-clauses exist, 
a RH choice is performed, but fact will not be relevant in
our statistical analysis: criticality of the cascade process will
always arise after a time interval throughout which an extensive number of
short-length clauses have been present (except at $X=\mathcal{O}(1)$).

If at some time $X$, $p_T$ variables are set
to True and $p_F$ to False in expectation, then $C_{i}$ changes
accordingly. If an
$i$-clause contains the variable just fixed, it is reduced to an
$(i-1)$-clause, and similarly an $(i+1)$-clause can be reduced to an
$i$-clause.  Call $\boldsymbol{\epsilon} = ( \epsilon, 1-\epsilon )$,
$\boldsymbol{p} = ( p_T , p_F )$, and $\boldsymbol{1} = (1 , 1 )$.
Still in expectation:
\be
  C_i(X+\boldsymbol{1} \cdot \boldsymbol{p}) = \left( C_{i}(X)
 - a \delta_{i,2} (1-\delta_{C_2(X),0})\right)
\left( 1 - \frac{i}{N-X} (\boldsymbol{1} \cdot \boldsymbol{p}) \,
  \right)
 + \frac{i+1}{N-X}
  (\boldsymbol{\epsilon} \cdot \boldsymbol{p}) \, C_{i+1}(X)
\ef;
\label{Cidyn}
\ee
where $a=0$ or 1 respectively in the case of RH$[p]$ and SCH. The two
heuristics are distinguished in that to initiate the cascades for
RH$[p]$ we have $\boldsymbol{p}_{\textrm{RH}[p]} = (p,1-p)$, and
for SCH we have $\boldsymbol{p}_{\textrm{SCH}} = (1,1)$. The SCH value
can be understood since setting one random literal in a two clause
implies setting the other to the opposite value, thus setting
variables to either $\pm 1$ is equally likely in expectation.

A round can be described by incorporating the variables set in the
forced steps. Suppose that during a subcritical round $m_T$ variables
are set to True and $m_F$ to False in expectation (including the free
step), and call $\boldsymbol{m}$ the vector $(m_T,m_F)$.
To leading order in $N-X$ the variation is
\be
  C_i(X+
\boldsymbol{1} \cdot \boldsymbol{m}
) = \left( C_{i}(X)
 - a \delta_{i,2} (1-\delta_{C_2(X),0})\right)
\left( 1 - \frac{i}{N-X} (\boldsymbol{1} \cdot \boldsymbol{m}) \,
  \right)
 + \frac{i+1}{N-X} 
  (\boldsymbol{\epsilon} \cdot \boldsymbol{m}) \, C_{i+1}(X)
\ef,
\label{Ciround}
\ee
A final simplification in the clause dynamics is to summarize the
behavior by continuous variables $x=X/N$ and $c_i=C_i/N$. In the
hypothesis of subcriticality, $\boldsymbol{m}/N$ is infinitesimal,
and we attain a differential equation description
\be
  \frac{\mathrm{d}}{\mathrm{d}x} c_i(x)
=
- a \delta_{i,2} \theta(c_2(x))
\eval{
\frac{1}{\boldsymbol{1} \cdot \boldsymbol{m}} }
+ \frac{1}{1-x}
\left(- i c_i(x) + (i+1) 
\eval{
\frac{\boldsymbol{\epsilon} \cdot \boldsymbol{m}} 
 {\boldsymbol{1} \cdot \boldsymbol{m}} }
c_{i+1}(x) \right)
\ef, 
\label{ciroundSC}
\ee
For both RH$[p]$ and SCH rules, the equation for
$c_3(x)$ gives
\be
c_3(x) = \frac{\gamma}{3} (1-x)^k
\ef,
\label{eq.c_k}
\ee
Instead, for $c_2(x)$ the equation is
nonlinear.
Indeed we get that
$\eval{m_T}$ and $\eval{m_F}$ are given by the combination of
equations (\ref{geometric_series}, \ref{eq.calM}, \ref{eq.c_k}), and
thus depend on the unknown function $c_2(x)$ (besides, of course, $x$,
$\epsilon$ and $\gamma$).
Using this expression within (\ref{ciroundSC}) allows us to determine
$c_2(x)$ by numerical integration, and thence
$\lambda_{\textrm{max}}(x)$.

The best choice for the parameter $p$ in RH$[p]$ is the one which
creates the smallest cascades, i.e.~the one ``more orthogonal'' to
the principal eigenvalue
$\boldsymbol{\epsilon}=(\epsilon, 1-\epsilon)$, but compatible
with the probabilistic interpretation of $\boldsymbol{p}$. Thus, in
the whole interval $\epsilon \in [0,1/2]$, the choice $p=1$ is
optimal.

Here we thus show the results for RH$[1]$ and for SCH. The latter is
always at least as good as the former, and gives a lower bound of
$\gamma_{\rm sch} = 1.6393$,
while RH$[1]$ attains
$\gamma_{\rm rh} = 1.6031$, for the case
$\epsilon=0$. 
Kalapala and Moore calculated these quantities
for positive 1-in-$K$ SAT, with compatible results for the $K=3$
case (up to maybe a misprint exchanging RH$[p]$ with RH$[1-p]$).

\begin{figure}
\caption{\label{fig.lambdas}
On the left, profiles of $\lambda(x)$ along decimation time $x$, for
RH$[1]$, at various $\epsilon$ and at the corresponding critical value
of $\gamma$. In all the cases, the functions $\lambda(x)$ are concave
(up to the limit value $\epsilon=1/2$, where $\lambda(x)=1-x$). For
$\epsilon$ larger or smaller than the tricritical value
$\epsilon^* = 0.272633$, the maximum of $\lambda(x)$ is achieved
respectively at $x=0$ or at $x>0$.
On the right, critical curves 
$\gamma_{\rm sch}(\epsilon)$
and $\gamma_{\textrm{RH$[1]$}}(\epsilon)$, obtained through
short-clause (SCH) and random heuristic at optimal parameter $p=1$
(RH$[1]$).
For a comparison with other values and curves here out of range,
cfr.~figures
\ref{fig_first}, and \ref{fig_RS}.}
\begin{center}
\setlength{\unitlength}{50pt}
\begin{picture}(5,2.7)(0,-0.1)
\put(-0.5,-0.2){\includegraphics[scale=0.4, bb=0 0 270 250]
    {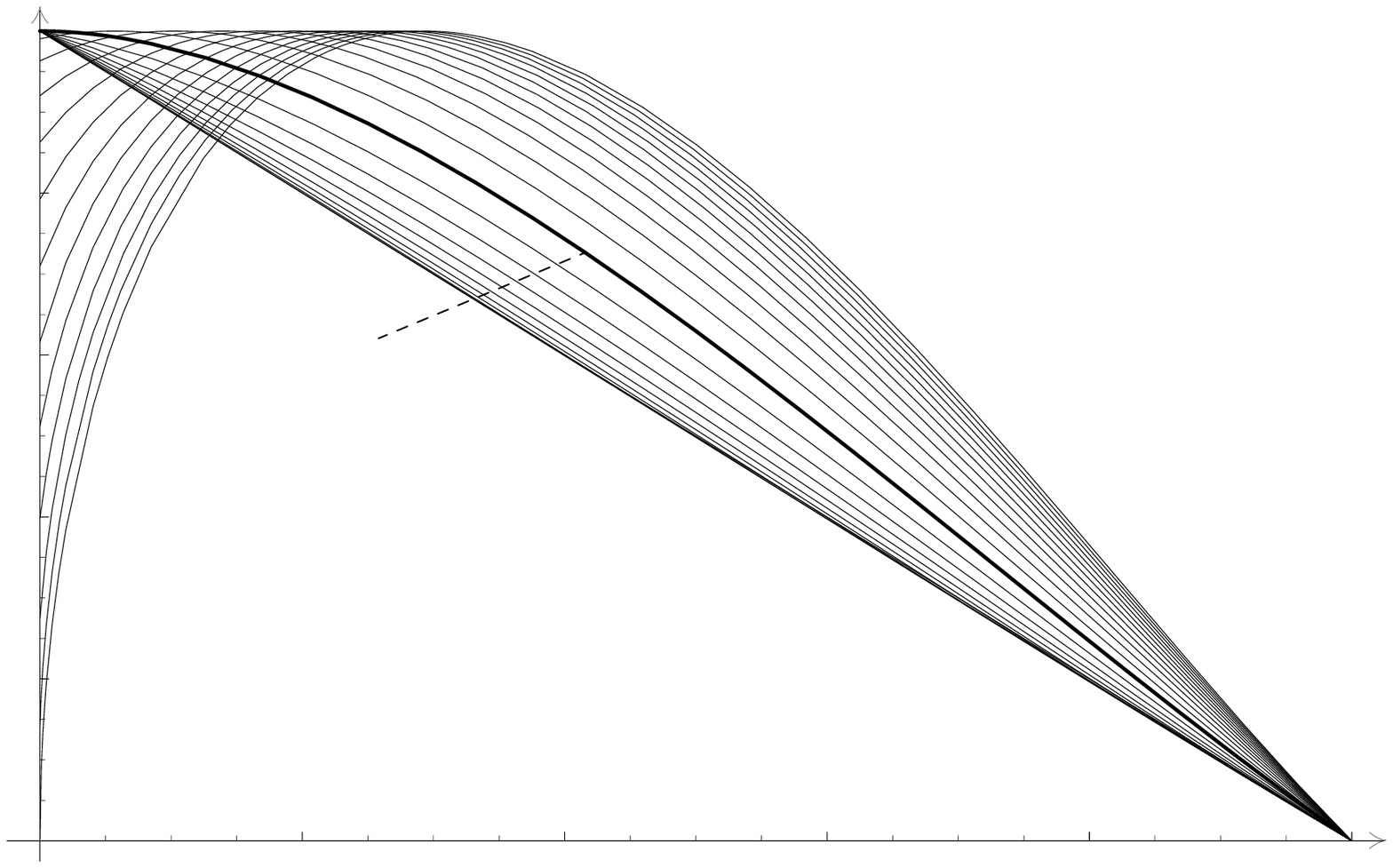}}
\put(4.35,-0.15){$x$}
\put(0,2.45){$\lambda(x)$}
\put(2.5,1.9){$\epsilon=0$}
\put(1.55,1.05){$\epsilon=0.5$}
\put(1.05,1.25){$\epsilon=\epsilon^*$}
\put(0.3,0.350){\makebox[0pt][r]{$0.2$}}
\put(0.3,0.816){\makebox[0pt][r]{$0.4$}}
\put(0.3,1,282){\makebox[0pt][r]{$0.6$}}
\put(0.3,1.748){\makebox[0pt][r]{$0.8$}}
\put(0.3,2.214){\makebox[0pt][r]{$1  $}}
\put(1.180,-0.23){\makebox[0pt][c]{$0.2$}}
\put(1.932,-0.23){\makebox[0pt][c]{$0.4$}}
\put(2.684,-0.23){\makebox[0pt][c]{$0.6$}}
\put(3.436,-0.23){\makebox[0pt][c]{$0.8$}}
\put(4.188,-0.23){\makebox[0pt][c]{$1  $}}
\end{picture}
\begin{picture}(3.8,2.7)(0,0.15)
\put(-0.1,0.05){\includegraphics[scale=1., bb=0 0 270 250]
    {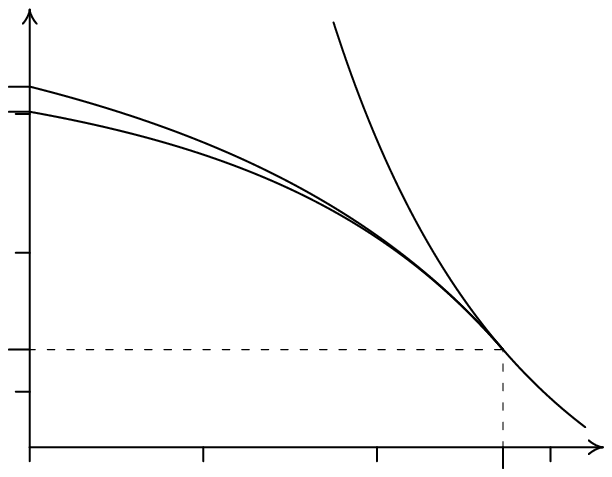}}
\put(0.05,0){$0$}
\put(1,0){$0.1$}
\put(2,0){$0.2$}
\put(3,0){$0.3$}
\put(2.45,-0.15){$0.272633$}
\put(3.53,0.2){\large $\epsilon$}
\put(-0.,2.7){\makebox[0pt][r]{\large $\gamma$}}
\put(-0.1,0.52){\makebox[0pt][r]{$1.2$}}
\put(-0.1,1.32){\makebox[0pt][r]{$1.4$}}
\put(-0.1,0.763){\makebox[0pt][r]{$1.261$}}
\put(-0.1,2.12){\makebox[0pt][r]{$1.603$}}
\put(-0.1,2.276){\makebox[0pt][r]{$1.639$}}
\put(2.3,1.9){$\gamma_{\rm uc}=\frac{1}{4 \epsilon (1-\epsilon)}$}
\put(1,2.15){$\gamma_{\rm sch}$}
\put(1,1.7){$\gamma_{\textrm{rh$[1]$}}$}
\end{picture}
\end{center}
\end{figure}

\subsection{Exact SAT/UNSAT threshold for $\epsilon>0.2726$}
\label{ssec.ESU}

This section proves the coincidence of the curves
$\gamma_{\rm sch}(\epsilon)$ and
$\gamma_{\rm uc}(\epsilon)$ for $\epsilon>0.2726$.
It was shown in the previous section that whenever
the cascades remain subcritical we are in the Easy-SAT phase. 
The criteria for the cascades to be subcritical at $x=0$ is precisely
$\gamma<\gamma_{\rm uc}(\epsilon)$.
It thus suffices to show 
that the maximum (over the
decimation time $x$) of the $\max_i |\lambda_i(x)|$, is attained for
$x=0$. This is indeed what happens in the interval $\epsilon \in [0.2726,1/2]$.

Building on the previous section we will see that, for
$\epsilon$--1-in-3-SAT and our heuristics,
$\lambda(x)$ is a concave function. So, the interval on which
$\gamma_{\rm uc}(\epsilon)$ and $\gamma_{\rm sch}(\epsilon)$ coincide
is the one in which
\be
\left.
\frac{\mathrm{d} \lambda(x; \epsilon, \gamma=\gamma_{\rm uc}(\epsilon))}
{\mathrm{d} x}
\right|_{x=0} \leq 0
\ef,
\label{localmaxx}
\ee
the endpoint being determined by the corresponding equality.

It is possible to calculate the characteristic polynomial (and
differentiate w.r.t.~$x$).
The expressions thereby found can, however, only be evaluated exactly at $x=0$.
At this value we have expressions for $c_i(x)$ and there derivatives
in terms of the initial conditions and $\boldsymbol{m}$.

As we increase connectivity towards the critical limit,
$4 \gamma \, \epsilon (1-\epsilon)=1$,
a further simplification is in the eigenvectors 
of ${\cal M}$ which become $\boldsymbol{\epsilon}$ and its orthogonal,
the latter having null eigenvalue.
Regardless of $\boldsymbol{p}$
(which must have a component along $\boldsymbol{\epsilon}$ in order to
have positive components), we have
\begin{align}
{\cal F}(x,\epsilon)
&:=
\frac{\boldsymbol{\epsilon} \cdot \boldsymbol{m}} 
 {\boldsymbol{1} \cdot \boldsymbol{m}}
\ef;
&
{\cal F}(0,\epsilon)
&=
\frac{\boldsymbol{\epsilon} \cdot \boldsymbol{\epsilon}} 
 {\boldsymbol{1} \cdot \boldsymbol{\epsilon}}
=1 - 2 \epsilon (1-\epsilon)
\ef.
\end{align}
Finally, the condition (\ref{localmaxx}) becomes
\be
\left( 1 + \frac{1}{4} 
\left(\frac{\epsilon}{1-\epsilon} 
    + \frac{1-\epsilon}{\epsilon}\right)^2 \right)
(1 - 2 \epsilon (1-\epsilon))
 - 2 \leq 0
\ef.
\label{criteria}
\ee
So that finally, after the change of variable 
$x=2 \epsilon (1-\epsilon)$, one gets the equation for the endpoint of
the interval
\be
2 x^3 - 2 x^2 + 3 x - 1 = 0\, ,
\ee
whose only real solution gives $\epsilon = 0.272633$, or its symmetric
point.

To show that the properties at $x=0$ are sufficient to determine
$\gamma_{\rm sch}$ it is necessary to show that whenever
criteria (\ref{criteria}) is met, and $\lambda(0)<1$, the algorithm is
subcritical at all $x$. If we want a true analytic proof, besides the
numerical evidence of figure \ref{fig.lambdas} (left), a method is
to find a function ${\hat \lambda}(x)$ such that
\be
\lambda(x) \leq {\hat \lambda}(x)\leq \lambda(0)
\ef,
\label{ubcond}
\ee
 hence establishing the result.

Since we find that $\lambda(x)$ is a monotonically increasing function of $c_2(x)$,
an upper bound ${\hat c}_2(x)>c_2(x)$ implies an upper bound in
$\lambda(x)$ also which we take to be ${\hat \lambda}(x)$. 
The bound function ${\hat c}_2$ is defined by replacing
the complicated function ${\cal F}(x,\epsilon)$ by the constant value
${\cal F}(0,\epsilon)$ in the expression
(\ref{ciroundSC}), which are then exactly solvable for all $x$ as
\be
{\hat c}_2(x)
= \gamma x (1-x)^2 {\cal F}(0,\epsilon)
= \gamma x (1-x)^2 \big( 1-2 \epsilon (1-\epsilon) \big)\, .
\label{chat}
\ee
For RH$[\frac{1}{2}]$ and certain other heuristics this approximation
can be shown to produce an upper bound for $c_2(x)$, and yet be exact
at $x=0$ in both absolute value and derivative.

This then allows an exact expression for 
$\frac{\mathrm{d} \hat{\lambda}(x)}{\mathrm{d} x}$ 
to be written in terms of $x$. 
Though the dependency on $x$ remains complicated it can be established that
\begin{align}
\frac{\mathrm{d} {\hat \lambda}(x)}{\mathrm{d} x} 
&< 0
&&\hbox{and}
&
{\hat \lambda}(0) &< 1
\ef,
\end{align}
exactly in the same interval of $\epsilon$ in which
(\ref{criteria}) holds. These fact proves that our ``local'' analysis
at $x=0$ was sufficient
for the purpose of identifying the maximum over $x$ of 
$\lambda(x)$ in this interval.

As a final remark we note that the proof of this exact bound is indirectly 
reliant on the convexity of the curves for all $\epsilon$ (figure \ref{fig.lambdas} left). 
Interestingly we found that for $\epsilon$--1-in-3-SAT with $k>3$ the curves 
are not convex for some $\epsilon$; 
the gradient at $x=0$ may be negative and yet the global 
maximum in lambda appears at $x>0$. On first inspection a rigorous 
bound appears more challenging to obtain in these cases.

\section{Other bounds}
\label{other_bounds}

\subsection{Upper bound from first moment method}
\label{app_1st}

Here we obtain the statistical properties of the 2-core of random
bipartite graphs, in the Erd\"os-R\'enyi ensemble described in section
\ref{model}, with $K=3$. Assuming $\gamma > 1/2$, the percolation
transition, we solve self-consistently for the probability that a
given branch of the graph is not percolating. 

We use ``giant'' or ``small'' for synonymous of ``of size
of order $N$'' or ``of size of order 1'' respectively.
Indeed, for graphs in our ensemble, a.s.~there is a single
giant 2-core component. Each edge is either attached on both sides to
a small tree; is attached to a small tree on one of the two
extrema, and to the giant 2-core on the other one (i.e.~it is in the
leaf-part of the giant component); or is connected to the 2-core
through both extrema. Only in this last case is it in the 2-core
of the graph.

Consider an incoming edge from a clause on the original graph. Call
$q$ the probability that the part of the graph ``downstream'' is a
tree. The edge will be attached to a variable, 
participating in $k$ other clauses 
($k$ Poissonian distributed of rate $\gamma$). For each clause there 
will be two incoming edges, which must also be
connected to finite trees.
Self-consistency will thus require that
\be
q= \sum_k e^{-\gamma} \frac{\gamma^k}{k!} q^{2 k} = e^{-\gamma(1-q^2)}
\ef.
\ee
The two functions $q(\gamma)$ and $\gamma(q)$, inverse of each other,
are both monotonic on our domains $q \in [0,1]$, 
$\gamma \in [1/2,+\infty)$. In particular, $\gamma(q)$ has an
algebraic form
\be
\gamma(q) = -\frac{\ln q}{1-q^2}
\ee
so that using $q$ as a parameter instead of $\gamma$ will simplify our
equations.

The probability that an incoming branch from a variable is connected
to a tree is $q^2$, as it is the probability that both outgoing branches
 from the neighbouring clause are connected to a tree. Thus, the
average number of variables of coordination $k \geq 2$, $\eval{N_k}$,
is proportional to a Poissonian with rate 
$\gamma (1-q^2) = - \ln q$.  Then, there will be 
$\eval{M_3} = (1-q)^3 M$ clauses remaining of degree 3, and
$\eval{M_2} = 3q(1-q)^2 M$ clauses reducing to 2-sat clauses, all the others
are decimated by the leaf removal. The number of edges is 
$E = \sum_k k N_k = 3 M_3 + 2 M_2$, so on average 
$\eval{E} = N (-\ln q) (1-q)$. All these averaged quantities are
concentrated.

Consider the ensemble of configurations $s$, in ``spin'' notations as
in the rest of the paper.
Call $x_k$ the fraction of variables of degree $k$ which take value
$+1$: the space of configurations is thus described by the infinite
vector $\{ x_k \}_{k \geq 2}$, with each $x$ in $[0,1]$, and a vector
$\{ x_k \}$ comes with an entropy
\be
S_{\textrm{var}}(\vec{x})
=\sum_k N_k h(x_k)\, ,
\ee
where we use the common two-state entropy function
$h(x)=-x \ln x - (1-x) \ln (1-x)$.
Denote by $x$ the fraction of incoming edges from variables assigned value $+1$
\be
x=\frac{1}{E} \sum_k k N_k x_k\, ,
\ee
and by $p$ the fraction of edges $(ai)$ such that 
$J_{ai} s_i = +1$, and hence 
$p=(1-\epsilon) x + \epsilon (1-x) =\epsilon + (1-2\epsilon) x$. 

The probability that a 1-in-3--clause
is satisfied is thus $3 p (1-p)^2$, and the probability that a 2-sat
clause is satisfied is $1-p^2$. So we get for the entropy term coming
from clauses
\be
S_{\textrm{cla}}(p(\vec{x}))
=M_3 \ln (3 p (1-p)^2) + M_2 \ln (1-p^2)\, .
\ee
The upper bound on the SAT/UNSAT transition is achieved by
the line, in the $(q,\epsilon)$ plane 
(with range $[0,1] \times [0,1/2]$), where the total (intensive) entropy
$S(q,\epsilon)$ vanishes:
\begin{align}
S(q,\epsilon)
&= 
\max_{ \{x_k\} }
S(q,\epsilon; \vec{x})
\ef;
&
S(q,\epsilon; \vec{x})
& =
S_{\textrm{var}}(\vec{x}) + S_{\textrm{cla}}(p(\vec{x}))
\ef.
\end{align}
This variational problem is infinite dimensional, thus at first sight
intractable.
Instead, stationarity with respect to~$x_k$ produces
\be
\label{eq.ydip}
\frac{1}{k} \ln \frac{x_k}{1-x_k}
=
(1-2 \epsilon) 
\left(
\frac{(1-q)}{3(1+q)} \frac{1-3p}{p(1-p)} 
- \frac{q}{1+q} \frac{2p}{1-p^2}
\right)
=: y(p)
\ef.
\ee
Remarkably, a single parameter $y$ describes the family of
possibly stationary vectors $\{ x_k \}$, this being a residue of the
original independence of the Poisson ensemble.
\be
\label{eq.874234}
x_k(y) = \frac{1}{1+e^{-ky}}
\ef.
\ee
Then, we can get self-consistently $p$ from $y$, through the $x_k$'s
\be
\label{eq.pdiy}
p(y)=
\epsilon + (1-2\epsilon)
\frac{1}{E} \sum_k k N_k
\frac{1}{1+e^{-ky}}
=
\epsilon + (1-2\epsilon)
\frac{q}{1-q} \sum_k \frac{(- \ln q)^k}{k!}
\frac{1}{1+e^{-(k+1)y}}
\ef.
\ee
So, the expression for the entropy $S(q, \epsilon)$
is given by the function
\be
S(q, \epsilon)
= q \sum_k \frac{(-\ln q)^k}{k!} h\left( \frac{1}{1+e^{ky}} \right)
+ 
\frac{(-\ln q)(1-q)}{3(1+q)}
\big(
(1-q) \ln (3p(1-p)^2) + 3q \ln (1-p^2)
\big)\, ,
\ee
where the values of $p$ and $y$ are determined by the (only) solution
of the nonlinear system of two equations (\ref{eq.ydip}) and
(\ref{eq.pdiy}).
The set of points $(\epsilon, \gamma(q))$ where the function $S(q,\epsilon)$
vanishes describes a curve which appears in the figures 
\ref{fig_first} and~\ref{fig_RS}.

Finally, we remark that a better upper bound can be achieved if one
realizes that a further removal procedure is allowed: if a variable is
connected only to 2-sat clauses, and with edges all of the same sign,
then one can safely fix it, satisfying all the neighboring
clauses. The new reduced instance is SAT if and only if the
original one is, but the number of solutions is potentially smaller:
as this decreases fluctuations, the bound is improved.

For the case $\epsilon=0$, this program is achieved in \cite{KSM04},
although in the restriction to the variational space of $x_k$ all
equal, and leads in that case to the (better) bound $\gamma_c \leq
1.932$.

\subsection{Algorithmic upper bound through embedding into 3-XOR-SAT}
\label{XOR-SAT}

An instance of 1-in-3-SAT is SAT only if the corresponding 3-XOR-SAT
instance is SAT, where 3-XOR clauses allows also for the extra
``spurious'' configuration 
$(J_1 \sigma_1, J_2 \sigma_2, J_3 \sigma_3) = (+,+,+)$.

Random Erd\"os-R\'enyi graphs with $K=3$ have a finite core (under
hypergraph leaf removal: if a variable has degree 1, one removes it
together with the incident clause) beyond a ``dynamical'' threshold
$\alpha_d = 0.818$. In a range $\alpha_d < \alpha < \alpha_c = 0.918$
there is an exponential number of solutions in the XOR-SAT problem,
even if restricted to the core. However, beyond the critical value
$\alpha_c$ there are no longer solutions (other than the single trivial
one, with all $\sigma_i=+1$, in the case
of ``positive'' instances with all $J_{ai}=+1$). These results can be
found, for example, in \cite{fede3xor, MRZ03}.

The last situation can be detected in polynomial time,
by Gaussian elimination on the adjacency matrix: if the rank
equals the number of variables (more generally, if it is smaller by at most
$\mathcal{O}(\ln N)$), the solutions on the core can be checked in
polynomial time. As a.s.~all the variables are forced to be
$+1$, a fraction of order 1 of the clauses in the core will be proven
to be satisfiable
only by the ``spurious'' configuration $(+,+,+)$. This provides a
certificate of unsatisfiability for the original instance in the
random $\epsilon$--1-in-3-SAT ensemble.

So, at all $\epsilon$, for $\gamma > 3 \alpha_c = 2.754$, one gets
a.s.~a certificate of unsatisfiability in randomized cubic
time (an upper bound for matrix triangulation). This proves that the
Easy-UNSAT phase starts from a finite $\gamma$ at all $\epsilon$.

The method strongly relies on the fact that a XOR-SAT core exists in
the instance. Unfortunately, this is not the case for the customary
reductions of SAT to $\epsilon$--1-in-$K$-SAT, even at large $\alpha$
(as the former constraints are much
more sloppy than the latter, the reduction makes use of auxiliary
leaf structures), so the method does not extend to a
randomized polynomial-time algorithm for finding a certificate in
large-$\alpha$ 3-SAT instances, in
agreement with the widespread conjecture that such an algorithm can
not exist~\cite{CSz}.

\section{1RSB at positive energy (finite $y$)}
\label{yfinite}

In this appendix we describe the 1RSB solution in the energetic 
zero-temperature limit, but at finite value of parameter $y$.
That allows us to compute the dependence of complexity 
$\Sigma$ on energy $E$ for a given probability of negation 
$\epsilon$ and connectivity~$\gamma$.

The survey propagation equations (\ref{SP_1}), (\ref{SP_2}) 
at finite $y$ becomes 
\begin{subequations}
\label{SP_1_y}
\begin{align}
p_+^{i \to a} + p_0^{i \to a}
&=
{\cal N}_{i \to a}^{-1} 
\Big( \sum_r \, e^{-yr} \, 
  \textrm{prob}(\Delta \tilde E^{i\to a}=r,
        \sum_{b\in \di \setminus a} u_{b\to i} \ge 0 )
\Big)
\ef;
\\
p_-^{i \to a} + p_0^{i \to a} 
&=
{\cal N}_{i \to a}^{-1}
\Big( \sum_r  \, e^{-yr} \,
  \textrm{prob}(\Delta \tilde E^{i\to a}=r,
        \sum_{b\in \di \setminus a} u_{b\to i} \le 0 )
\Big)
\ef;
\\
p_0^{i \to a} 
&=
{\cal N}_{i \to a}^{-1}
\Big( \sum_r \, e^{-yr} \,
  \textrm{prob}(\Delta \tilde E^{i\to a}=r, 
        \sum_{b\in \di \setminus a} u_{b\to i} = 0 ) 
\Big)
\ef;
\end{align}
\end{subequations}
where ${\cal N}_{i \to a}$ is the normalization factor, and
$\Delta \tilde E^{i\to a} =
 \sum_{b\in \di \setminus a} |u_{b\to i}|
 - \left|\sum_{b\in \di \setminus  a} u_{b\to i}\right|$.
When using Hamiltonian $\ham$ as in (\ref{eq.ham}) (as we did), the
probabilities of biases
are given by the incoming probabilities of fields as
\begin{subequations}
\label{SP_2_y}
\begin{align}
q_{J_{ai}}^{a \to i}
&=
{\cal N}_{a \to i}^{-1} \, 
  p^{j \to a}_{-J_{aj}} p^{k \to a}_{-J_{ak}}
\ef;
\\
\label{eq.sp2y_b}
q_{-J_{ai}}^{a \to i}
&= {\cal N}_{a \to i}^{-1} \, 
\big(
p_{J_{aj}}^{j \to a} (1-p_{J_{ak}}^{k \to a} )
 + (1-p_{J_{aj}}^{j \to a} )p_{J_{ak}}^{k \to a} \big)
\ef;
\\
\label{eq.sp2y_c}
q_{0}^{a \to i} 
&=
{\cal N}_{a \to i}^{-1} \, 
\big[ \big(
p_{-J_{aj}}^{j \to a} p_{0}^{k \to a}
 + p_{0}^{j \to a} p_{-J_{ak}}^{k \to a}
 + p_{0}^{j \to a} p_{0}^{k \to a} \big)
      + p^{j \to a}_{J_{aj}} p^{k \to a}_{J_{ak}} \, e^{-y}
\big]
\ef;
\end{align}
\end{subequations}
where ${\cal N}_{a \to i}$ is again the normalization.  If we were
using $\ham'$ of equation (\ref{eq.ham'}) instead, the last summand in
(\ref{eq.sp2y_c}), proportional to $e^{-y}$, would have appeared
in~(\ref{eq.sp2y_b}).

Similarly, for the replicated free energy and energy we have
\begin{align}
-y\Phi(y) 
&=
\sum_a \ln \Big( 
\sum_r \, e^{-yr} \, \textrm{prob}(\Delta E^{a \cup \da}=r)
      \Big)
- \sum_i (d_i-1) \ln \Big(
\sum_r \, e^{-yr} \, \textrm{prob}(\Delta E^{i}=r)   
      \Big)
\ef;
\label{Phi_y}
\\
E(y)
&=
\sum_a \frac{ \sum_r r\, e^{-yr} \, \textrm{prob}(\Delta E^{a \cup \da}=r)}
     {\sum_r \, e^{-yr} \, \textrm{prob}(\Delta E^{a \cup \da}=r)}
- \sum_i (d_i-1) \frac{ \sum_r r \, e^{-yr} \, \textrm{prob}(\Delta E^{i}=r)}
     { \sum_r\, e^{-yr} \, \textrm{prob}(\Delta E^{i}=r)}
\ef;
\label{E_y}
\end{align}
The probabilities $\textrm{prob}(\Delta E=r)$ have to be computed 
algorithmically (sum over all combinations). We found a closed 
formulas only when $r=0$ (no contradictions), 
equations (\ref{SP_1}) and~(\ref{eq.prob_12}).

\subsection{Complexity as a function of energy}
\label{1RSBcomp}

\begin{figure}[!ht]
\setlength{\unitlength}{40pt}
\begin{picture}(10,6.7)
  \put(0,0){\includegraphics[scale=1]{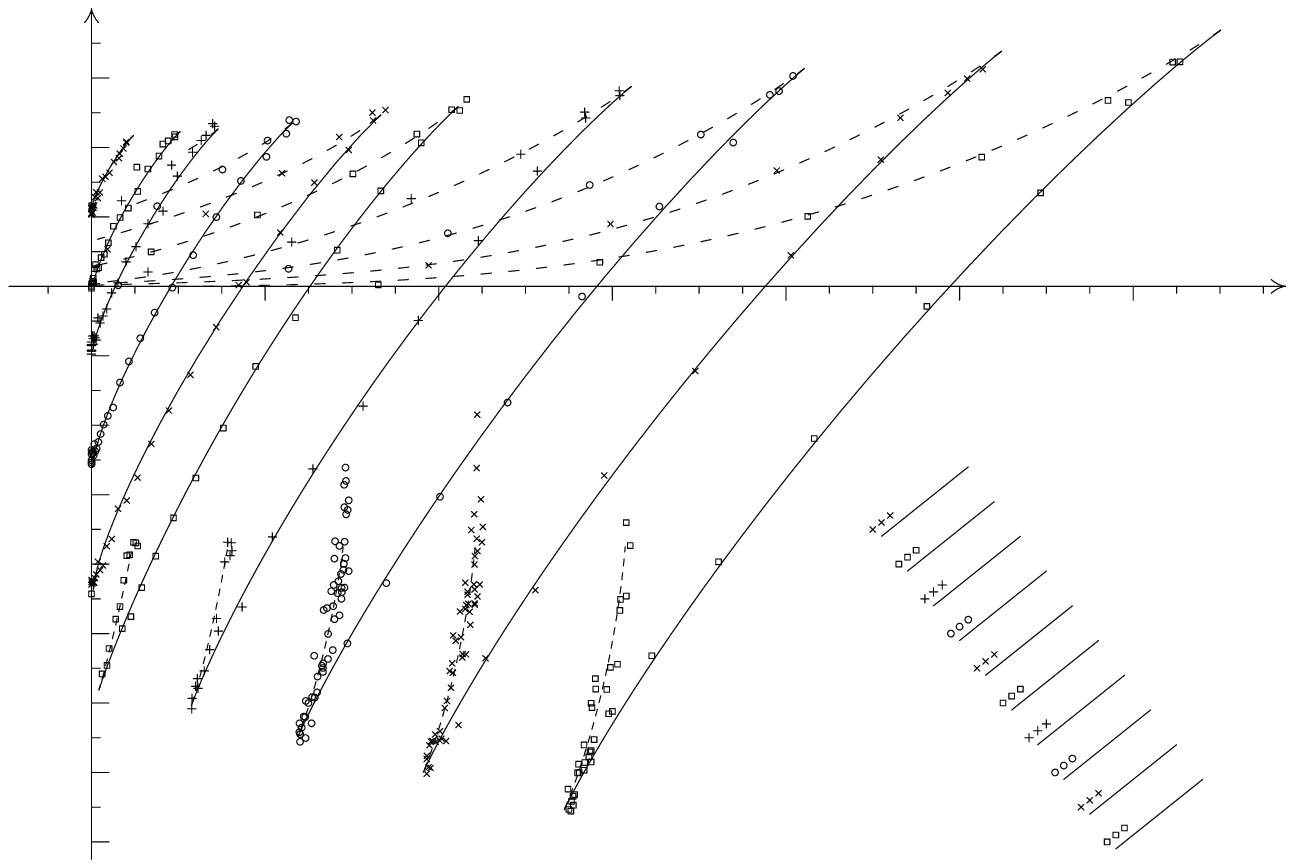}}
  \put(8.8,4.05){$\times 10^{-2}\ E/N$}
  \put(-0.35,6.35){$\frac{1}{N} \Sigma(E/N)$}
\put(6.9 , 3.4){For curves from left to right:}
  \put(7.25 , 3.1 ){$\scriptstyle{\gamma = 1.850}$}
  \put(7.4375, 2.85 ){$\scriptstyle{\gamma = 1.879 \, \simeq \, \gamma^*}$}
  \put(7.625 , 2.6 ){$\scriptstyle{\gamma = 1.900}$}
  \put(7.8125, 2.35 ){$\scriptstyle{\gamma = 1.935}$}
  \put(8. , 2.1 ){$\scriptstyle{\gamma = 1.970}$}
  \put(8.1875, 1.85 ){$\scriptstyle{\gamma = 2.000}$}
  \put(8.375 , 1.6 ){$\scriptstyle{\gamma = 2.050}$}
  \put(8.5625, 1.35 ){$\scriptstyle{\gamma = 2.100}$}
  \put(8.75 , 1.1 ){$\scriptstyle{\gamma = 2.150}$}
  \put(8.9375, 0.85 ){$\scriptstyle{\gamma = 2.200}$}
  \put(8.6, 2.){$\scriptstyle{\cdots\cdots\ (\gamma_p = 1.992)}$}
  \put(0.675,4.135){$0$}
  \put(2.00,4.15){$\scriptstyle{0.2}$}
  \put(3.29,4.15){$\scriptstyle{0.4}$}
  \put(4.51,4.15){$\scriptstyle{0.6}$}
  \put(5.76,4.15){$\scriptstyle{0.8}$}
  \put(7.01,4.15){$\scriptstyle{1.0}$}
  \put(8.26,4.15){$\scriptstyle{1.2}$}
  \put(0.8 ,4.85){\makebox[0pt][r]{$\scriptstyle{0.002}$}}
  \put(0.8 ,5.35){\makebox[0pt][r]{$\scriptstyle{0.004}$}}
  \put(0.8 ,5.85){\makebox[0pt][r]{$\scriptstyle{0.006}$}}
  \put(0.8 ,3.85){\makebox[0pt][r]{$\scriptstyle{-0.002}$}}
  \put(0.8 ,3.35){\makebox[0pt][r]{$\scriptstyle{-0.004}$}}
  \put(0.8 ,2.85){\makebox[0pt][r]{$\scriptstyle{-0.006}$}}
  \put(0.8 ,2.35){\makebox[0pt][r]{$\scriptstyle{-0.008}$}}
  \put(0.8 ,1.85){\makebox[0pt][r]{$\scriptstyle{-0.010}$}}
  \put(0.8 ,1.35){\makebox[0pt][r]{$\scriptstyle{-0.012}$}}
  \put(0.8 ,0.85){\makebox[0pt][r]{$\scriptstyle{-0.014}$}}
  \put(0.8 ,0.35){\makebox[0pt][r]{$\scriptstyle{-0.016}$}}
\end{picture}
\caption{\label{sigm_ener}
Complexity as a function of energy for positive 1-in-3 SAT, for
several different connectivities.
$\gamma=1.850$ is in the SAT region, $\gamma=1.879$ is near the SAT/UNSAT
transition, $1.900\le \gamma \le 1.970$ is in
the UNSAT region and $2.000\le \gamma \le 2.200$ is in the UNSAT region
where a solution at zero energy does not exists.}
\end{figure}

Again with the population-dynamics algorithm we solve 
eqs.~(\ref{SP_1_y})-(\ref{E_y}) and from the solution obtain the replicated 
potential $\Phi(y)$ (\ref{Phi_y}). The function $\Sigma(E)$, plotted
in fig.~\ref{sigm_ener}, is obtained by fitting function $\Phi(y)$ 
and computing the Legendre transform (\ref{Legr}) of the fit:
\begin{align}
E(y)      &=     \frac{\partial \big( y \Phi(y) \big) }{\partial y} 
\ef;
&
\Sigma(y) &= y^2 \frac{\partial  \Phi(y) }{\partial y}
\ef.
\end{align}
A non-paramagnetic solution of (\ref{SP_1_y},\ref{SP_2_y}) exists only above 
a value $y_{\textrm{ex}}(\gamma)$, 
which decreases monotonically, from its asymptotics 
$y_{\textrm{ex}} \to +\infty$ for $\gamma \searrow \gamma_{\rm sp}$ to
$y_{\textrm{ex}} \to 0$ for $\gamma \to \infty$.
The function $\Sigma(E)$ is
parametrically identified from the two $\Sigma(y)$ and $E(y)$ above. Assuming the
latter are regular functions and noting that 
$\partial \Sigma(E)/ \partial E = y$, one also finds that $\Sigma(E)$
is regular, and convex or concave respecively if parameter $y$ grows towards
right or left, up to (possibly) special points where the curve changes
concavity (\emph{turning points}).
Note that only the concave parts have a physical meaning, while the
convex parts are the ``non-physical'' portion of the formal solution.

\begin{table}
\[
\begin{array}{|c||c||c|c|c||c|c||c|c|c|}
\hline
\gamma & y_{\rm ex}
             & y_{\rm max}
                     & E_{\rm max}
                                & \Sigma_{\rm max}
                                          & y_{\rm gs}
                                                 & E_{\rm gs} & y_{\rm min}
                                                                     & E_{\rm min}
                                                                                & \Sigma_{\rm min} \\
\hline
1.850  & 2.9 & 3.35  & 0.000410 & 0.00414 &      & 0          &      &          & (+0.00247) \\
1.879  & 2.1 & 2.83  & 0.000969 & 0.00441 &      & 0          &      &          & (-0.00015) \\
1.900  & 1.6 & 2.64  & 0.00143  & 0.00471 & 5.31 & 0.000282   &      &          & (-0.00215) \\
1.935  & 1.0 & 2.40  & 0.00236  & 0.00480 & 4.31 & 0.000939   &      &          & (-0.00570) \\
1.970  & 0.6 & 2.25  & 0.00336  & 0.00517 & 3.83 & 0.00176    &      &          & (-0.00954) \\
2.000  & 0.4 & 2.09  & 0.00428  & 0.00534 & 3.60 & 0.00255    & 7.37 & 0.000149 & -0.0111 \\
2.050  & 0.1 & 2.04  & 0.00573  & 0.00611 & 3.27 & 0.00406    & 6.37 & 0.00107  & -0.0121 \\
2.100  & 0.1 & 1.94  & 0.00613  & 0.00815 & 3.04 & 0.00580    & 5.76 & 0.00232  & -0.0130 \\
2.150  & 0.1 & 1.86  & 0.00766  & 0.0104  & 2.88 & 0.00632    & 5.27 & 0.00377  & -0.0138 \\
2.200  & 0.1 & 1.67  & 0.00993  & 0.0126  & 2.72 & 0.00674    & 5.32 & 0.00549  & -0.0155 \\
\hline
\end{array}
\]
\caption{\label{y_maxmin} Values at special points in the complexity vs.~energy curves
 of figure \ref{sigm_ener}. The parametrizations $E(y)$ and
 $\Sigma(y)$ at various gamma exist for $y \gtrsim y_{\rm ex}(\gamma)$
 (the estimate is an upper bound, being the first value for which a
 non-paramagnetic RS solution emerged).
 All energies $E$ and complexities $\Sigma$ are extensive, and
 factors $1/N$ in the table entries are understood.
 The triplets $(y,E,\Sigma)_{\rm max}$ and
 (if any) $(y,E,\Sigma)_{\rm min}$ are the two turning points, while
 $(y,E,\Sigma=0)_{\rm gs}$ is the point where the complexity vanishes,
 along the physical branch of the curve. In the column for
 $\Sigma_{\rm min}$, the first 5 values are instead $\Sigma(E=0)$, as
 $E_{\rm min}=0$ without turning point in that case.}
\end{table}

\begin{figure}[!ht]
\setlength{\unitlength}{40pt}
\begin{picture}(6.5, 4.1)
  \put(0,0){\includegraphics[scale=0.5]{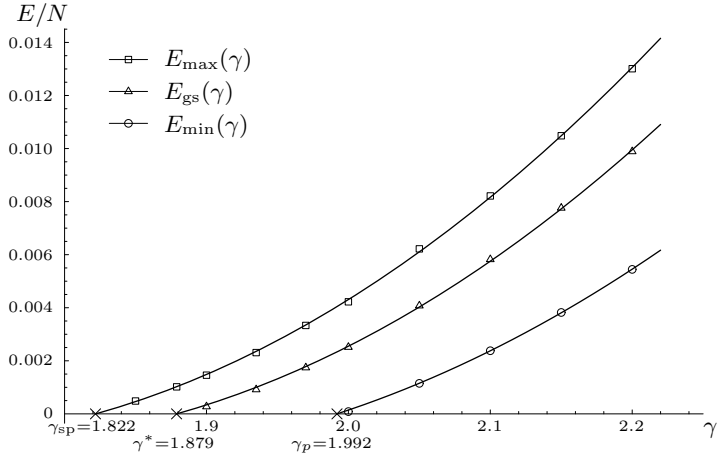}}
 \put(0,4.){$E/N$}
  \put(6.5,0.1){$\gamma$}
  \put(0.35 ,0.24){\makebox[0pt][r]{$\scriptstyle{0}$}}
  \put(0.35 ,0.74){\makebox[0pt][r]{$\scriptstyle{0.002}$}}
  \put(0.35 ,1.24){\makebox[0pt][r]{$\scriptstyle{0.004}$}}
  \put(0.35 ,1.74){\makebox[0pt][r]{$\scriptstyle{0.006}$}}
  \put(0.35 ,2.24){\makebox[0pt][r]{$\scriptstyle{0.008}$}}
  \put(0.35 ,2.74){\makebox[0pt][r]{$\scriptstyle{0.010}$}}
  \put(0.35 ,3.24){\makebox[0pt][r]{$\scriptstyle{0.012}$}}
  \put(0.35 ,3.74){\makebox[0pt][r]{$\scriptstyle{0.014}$}}
  \put(1.68 ,0.12){$\scriptstyle{1.9}$}
  \put(3.02 ,0.12){$\scriptstyle{2.0}$}
  \put(4.36 ,0.12){$\scriptstyle{2.1}$}
  \put(5.70 ,0.12){$\scriptstyle{2.2}$}
  \put(0.3 ,0.12){$\scriptstyle{\gamma_{\rm sp}=1.822}$}
  \put(1.1 ,-0.04){$\scriptstyle{\gamma^*=1.879}$}
  \put(2.6 ,-0.04){$\scriptstyle{\gamma_{p}=1.992}$}
  \put(1.4 ,3.57){$E_{\rm max}(\gamma)$}
  \put(1.4 ,3.26){$E_{\rm gs}(\gamma)$}
  \put(1.4 ,2.95){$E_{\rm min}(\gamma)$}
\end{picture}
\caption{\label{fig_energamma}
The values of $E_{\rm max}(\gamma)$, $E_{\rm gs}(\gamma)$
and $E_{\rm min}(\gamma)$, and a quadratic fit which extrapolates to
$E_{\rm max}(\gamma_{\rm sp}) = E_{\rm gs}(\gamma^*)= E_{\rm min}(\gamma_p) = 0$.}
\end{figure}  

It turns out that, for $\gamma_{\rm sp} < \gamma < \gamma_p$, there is a
single turning point, from convex at small $y$ to concave at large
$y$, the values $E_{\rm max}$ and $y_{\rm max}$ labeling the values of
$E$ and $y$ at this point. Instead, for $\gamma > \gamma_p$, a second
turning point, from concave to convex, arises at higher values of $y$,
thus defining the values $E_{\rm min}$ and $y_{\rm min}$.

The labels of ``max'' and ``min'' stand for the fact that there
exist phases with energy $E$ (a number approximatively
$\exp(\Sigma(E))$, if $\Sigma(E) > 0$, or in a fraction of instances
of order $\exp(\Sigma(E))$, if $\Sigma(E) < 0$), only for energies in
a range $E_{\rm min} < E < E_{\rm max}$ (where $E_{\rm min}=0$ for
$\gamma < \gamma_p$), while we should interpret that there are no pure
phases with energy $E$ out of the range above, up to a subexponential
fraction of instances.

We find that, for all $\gamma$, $\Sigma(E_{\rm max}) > 0$, while
for $\gamma > \gamma_p$, $\Sigma(E_{\rm min}) < 0$.
An intermediate value $E_{\rm min} < E_{\rm gs} < E_{\rm max}$,
corresponding to the one at which the complexity vanishes, 
exists for all $\gamma > \gamma^*$. It is
the value of minimum energy of a configuration in a typical instance
sampled from the corresponding ensemble, so it is important for the
statistical properties of the ``optimization'' problem in the UNSAT
phase.

In the table~\ref{y_maxmin} we show the numerical
values of the quantities described above, for a range of relevant
$\gamma$'s, obtained by population dynamics.

Let us mention that, in our interpretation, the possibility of hard
contradictions, peculiar to 1-in-3 SAT and other highly constrained 
NP-complete problems and absent in the more intensively studied $K$-SAT and
Coloring, is responsible also for the existence of the second turning
point, and the second unphysical branch at high values of $y$, which
is indeed a new feature of this system.

\section{Stability of 1RSB}
\label{S1RSB}

In this appendix we describe how to check the self-consistency
(stability) of the 1RSB solution, with a treatment
similar to the one in section \ref{RS_stab}, for the replica-symmetric
solution. We do it only for the solution at zero energy, 
$y \to \infty$: as we will see, this is sufficient to determine the
SAT/UNSAT transition line in an interval of $\epsilon$ near
$\epsilon=0$, thus complementing the informations we already have for
the neighbourhood of $\epsilon=1/2$.

The stability analysis of the replica-symmetric solution investigates
if the replica-symmetric state tends to split into exponentially many
states. In the case of 1RSB we have two stability conditions to test
this. The type I stability condition determines the tendency of 1RSB
states to aggregate, and the type II determines the tendency of the states
to split. The names type I and type II comes form
\cite{MR03,MPR03}.  In the case that the 1RSB solution is not stable,
i.e.~the states tend to split or aggregate, we would deduce
instability towards the 2-step of Replica-Symmetry-Breaking
(2RSB). However, this is not expected to be the correct picture, while
what more probably occurs is what is called full-RSB \cite{Parisi80},
or something else still unknown.

The method for computing these stabilities for models on random graphs
was introduced in \cite{MR03} and applied to $K$-SAT \cite{MPR03,MMZ05} and 
later to many other problems. For both types of 1RSB stability 
there exist several equivalent analyses. 

For the stability of type II we choose the 
{\it bug proliferation}. This is developed concisely in 
section \ref{S1RSBII}
for the 1RSB solution of zero energy ($y \to \infty$). For more
theoretical background for this method see
papers~\cite{MR03,RBMM04,MMZ05}.

For type I instability it is possible to consider the convergence of survey 
propagation equations (\ref{SP_1}, \ref{SP_2}) on a single graph. 
However, we choose to consider the {\it noise propagation} method,
which uses a population-dynamics technique similar to that used for 
stability of the replica-symmetric solution, eq.~(\ref{RS_MT}).  
Again we give just the formulas and results in section \ref{S1RSBI},
for general explanation see~\cite{MR03,RBMM04,MMZ05}.

\subsection{Stability of the second kind, bug proliferation}
\label{S1RSBII}

Suppose that, in a neighbourhood of edge $(a,i)$, with $j$ and $k$
being the other two variables incident on $a$, an incoming warning
$u_{b\to j}$ is changed from value $u$ to another value $u'$.  Assume
also that there are $n-1$ other clauses $c_2, \ldots, c_n$ (besides
$a$ and $b$) incoming to node $j$, the warnings having values
respectively $u_2, \ldots, u_n$, and that there are $m$ other clauses
$c'_1, \ldots, c'_m$ (besides $a$) incoming to node $k$, the warnings
having values respectively $v_1 \dots v_m$.  Conditioned to the
existence of the path $\big((bj),(ja),(ai)\big)$, the other
coordinations $m$ and $n-1$ are decorrelated and Poissonian
distributed, with rate $\gamma$.  Denote with the letter ${\cal J}$
the set of parameters describing the characteristics of the graph in
this neighbourhood
\be
{\cal J} = (n-1, m; J_{ai}, J_{aj}, J_{ak})
\ef;
\ee
Labels $w$ and $w'$ will describe the value of the output warning,
$u_{a \to i}$, under $u_{b \to j} = u$ and $u'$ respectively:
\[
\setlength{\unitlength}{50pt}
\begin{picture}(3,2.2)(0,0.1)
\put(0,0){\includegraphics[scale=1, bb=25 310 175 420, clip=true]
  {fig_smallFG.eps}}
\put(2.7,2.){$i$}
\put(1.85,2.){$a$}
\put(1,2.){$j$}
\put(0.2,2){$b$}

\put(1.95,2.15){$w(\to w')$}
\put(0.3,2.15){$u(\to u')$}

\put(0.35,1.05){$c_2$}
\put(0.95,0.9){$c_n$}

\put(1.0,0.35){$c'_1$}
\put(1.6,0.2){$c'_m$}
\end{picture}
\]
Define the six-dimensional transition matrix, in the index pairs $(u,u')$ and
$(w,w')$, pairs of distinct elements in
$\{0, \pm 1\}^2$
\begin{multline}
{\cal P}_{\cal J} (w \to w'|u \to u')= 
\frac{1}{\cal N}
\sum_{\substack{
u_2 \cdots u_n \in \{0,\pm 1\} \\ v_1 \cdots v_m \in \{0,\pm 1\} }}
q_{u_2}^{c_2 \to j}
\cdots 
q_{u_n}^{c_n \to j}
\,
q_{v_1}^{c'_1 \to k}
\cdots 
q_{v_m}^{c'_m \to k}
\,
\\
\cdot \,
\delta_{{\cal F}^{a\to i}(u,u_2 \dots,v_m),  w}  \;
\delta_{{\cal F}^{a\to i}(u',u_2 \dots,v_m),  w'} \;
\delta_{\Delta E^{a\to i}(u',u_2 \dots,v_m), 0}
\ef.
\end{multline}
The quantity ${\cal P}(w \to w'|u \to u')$ is 
proportional to the probability that the change $u \to u'$
in warning $u_{b\to j}$ has induced a change $w \to w'$ in 
the warning $u_{a\to i}$.
Here
${\cal F}^{a\to i}(u^{b \to j}, u^{c_2 \to j} \dots, u^{c_n \to j}, 
u^{c'_1 \to k} \dots, u^{c'_m \to k})$
implements the cavity equations at zero energy (\ref{eqs.cavityZT}),
with the appropriate
disorder parameters  $(J_{ai}, J_{aj}, J_{ak})$, so the delta's force
this value to be equal to $u^{a \to i}$, in the two cases
$(u^{b \to j}=u, u^{a \to i}=w)$ and 
$(u^{b \to j}=u', u^{a \to i}=w')$. The delta in the energy shift
is the residual of the reweightening factor 
$\exp( -y \Delta E^{a\to i})$ in the limit 
$y \to \infty$ (note that it only appears on the pair $(u', w')$ of
biases along the chain).
Normalizations ${\cal N}={\cal N}_{j\to a}
{\cal N}_{k\to a} {\cal N}_{a\to i}$ are those from 
eqs.~(\ref{SP_1}, \ref{SP_2}). 

The transition matrix defined above determines if a small fluctuation
in the equilibrium distribution of the fields is reinforced through
the cavity iterations, thus leading to an instability.
After $d$ iterations, the modulus of the fluctuation changes on
average by some factor 
$(2 \gamma)^d {\rm Tr} 
\langle {\cal P}_{{\cal J}_1} \dots {\cal P}_{{\cal J}_d} \rangle$,
because there are on average $(2 \gamma)^d$ chains of length $d$
ending on a given edge, and the trace of a ``chain'' of transition
matrices estimates the influence of changing a bias at an edge at
distance $d$ upstream.
So we define the (finite-$d$ and $d \to \infty$) type-II stability
parameters
\begin{align}
   \mu_{\rm II}(d)
&= 
2 \gamma \;
  {\rm Tr} \langle {\cal P}_{{\cal J}_1} \dots {\cal P}_{{\cal J}_d}
  \rangle^{\frac{1}{d}}
\ef;
&
   \mu_{\rm II} 
&= 
\lim_{d \to \infty}
   \mu_{\rm II}(d)
\ef;
\label{mu_d}
\end{align}
where the average is over the connectivity distribution and the 
disorder in negations, i.e.~the parameters globally identified above with
the letter ${\cal J}$.
The various ${\cal J}$'s refers to the different segments of the
chain, and thus are independent.
Again, the stability condition reads
\be
   \mu_{\rm II}  < 1
\ef.
\ee
The matrix ${\cal P}$ for the $\epsilon$--1-in-3 SAT problem is 
six-dimensional, and we computed it for general 
realizations of negation and connectivities. It has a block-triangular 
form (a change $0\to \pm 1$ never induces changes other than 
$0\to \pm 1$ and a change $\pm 1 \to 0$ never induces $\pm 1 \to \mp 1$).
Moreover two of the three $2 \times 2$ blocks on the diagonal, $B$, are equal
and have elements always larger than those of the third block $B'$
\[
{\cal P}(w \to w'|u \to u') : \qquad
\begin{array}{lccc}
 & \scriptstyle{0 \to \pm 1} & \scriptstyle{\pm 1 \to 0} &
 \scriptstyle{\pm 1 \to \mp 1} \\
\cline{2-4}
\scriptstyle{0 \to \pm 1}     &
\multicolumn{1}{|c}{\raisebox{-5pt}{\rule{0pt}{16pt}}B} 
& \multicolumn{1}{|c}{0} & \multicolumn{1}{c|}{0} \\
\cline{3-3}
\scriptstyle{\pm 1 \to 0}     &
\multicolumn{1}{|c}{\raisebox{-5pt}{\rule{0pt}{16pt}}\ast} 
& B & \multicolumn{1}{|c|}{0} \\
\cline{4-4}
\scriptstyle{\pm 1 \to \mp 1} &
\multicolumn{1}{|c}{\raisebox{-5pt}{\rule{0pt}{16pt}}\ast} 
& \ast & \multicolumn{1}{c|}{B'} \\
\cline{2-4}
\end{array}
\]
So in the large-$d$ limit
we need to analyze only the $2 \times 2$ block $B_{\cal J}$ of elements,
propagating the `bug' $0 \leftrightarrow \pm 1$. 
For ${\cal J}=(n-1,m;J_{ai},J_{aj},J_{ak})$, the block
elements take the form
\begin{gather}
B=
\left[
\delta_{J_{aj},+1}
\begin{pmatrix}
1 & 0 \\
0 & 1
\end{pmatrix}
+
\delta_{J_{aj},-1}
\begin{pmatrix}
0 & 1 \\
1 & 0
\end{pmatrix}
\right]
\begin{pmatrix}
  0   & \alpha \\
\beta & 0
\end{pmatrix}
\left[
\delta_{J_{ai},+1}
\begin{pmatrix}
1 & 0 \\
0 & 1
\end{pmatrix}
+
\delta_{J_{ai},-1}
\begin{pmatrix}
0 & 1 \\
1 & 0
\end{pmatrix}
\right]
\ef;
\\
\alpha = \prod_{\ell=1}^m q^\ell_0 \prod_{j=2}^n q^j_0
\ef; \qquad \qquad \qquad
\beta  = \prod_{\ell=1}^m \big( q^\ell_0 + q^\ell_{-J_{ak}} \big)
\prod_{j=2}^n q^j_0
\ef;
\end{gather}
and triplets $(q^e_0, q^e_-, q^e_+)$ for different edges
are independently sampled from the stationary distribution.

Results for the type-II stability
are presented in figure~\ref{stabII}.

\subsection{Stability of the first kind, the noise propagation}
\label{S1RSBI}

In a similar way to bug proliferation, we write a sort of transfer
matrix $T$
\begin{align}
\left( T^{a\to i}_{b\to j} \right)_{\sigma, \tau} 
&=
\frac{\partial q_{\tau}^{a\to i}}{\partial q_{\sigma}^{b\to j}}
\ef,
&
\sigma, \tau 
&\in \{ +, - \}
\ef.
\end{align}
The dependence of $q^{a\to i}$ on $q^{b\to j}$ is given by the survey propagation 
equations (\ref{SP_1},\ref{SP_2}).

We perform a population-dynamics analysis, where to every edge in the
population is associated a triple for the surveys, $(q_-,q_0,q_+)$,
updated with the cavity equations (\ref{SP_1},\ref{SP_2}), and a pair of
{\it noise} parameters, ${\vec v}=(v_+,v_-)$, which are updated according to
\begin{equation}
 {\vec v^{a\to i}} = 
\sum_{b \in \djj \setminus a}  T^{a\to i}_{b\to j}  {\vec v^{b\to j}} +  
\sum_{c \in \dk \setminus a}   T^{a\to i}_{c\to k}  {\vec v^{c\to k}}
\ef.
\end{equation}
The motivation is to compute whether
a small change in the equilibrated incoming survey $q^{b\to j}$ is
dumped under cavity iterations.

The analysis goes in complete analogy with the one in section \ref{RS_stab}.
We initialize the noise parameters with an arbitrary random procedure,
and wait for equilibration of the distribution, up to a scaling
overall, $\| v \|_t :=\sum_{e} \left(  |v^{e}_+| + |v^{e}_-| \right)$
where $t$ denotes the iteration time, and the sum is over the
population.
The stability parameter is now, for some time $t$ larger than equilibration,
\be
\label{lambda_I}
\mu_{\rm I} = \eval{ \frac{\| v \|_{t+1}}{\| v \|_{t}} }
\ef,
\ee
and the stability condition is $\mu_{\rm I} < 1$.


\end{document}